\documentclass[12pt]{article}
\usepackage{graphicx,bm,amsmath,amssymb,latexsym,psfrag,multirow,color}
 
\allowdisplaybreaks
\def\muf{\mu}

\newcommand{\non}{\nonumber}

\def\rd{\mathrm{d}}

\def\ecm{E_{\mathrm{CM}}}
\def\qq{ {q \bar{q}} }
\DeclareMathOperator{\Li2}{Li_2}

\def\Mx{m_X}

\def\Mxt{m_X^2}

\def\x{x}

\def\rd{\mathrm{d}}

\def\ptmax{p_T^{\mathrm{max}}}

\definecolor{darkred}{rgb}{0.6,0.0,0.0}
\definecolor{darkblue}{rgb}{0.0,0.0,0.5}
\definecolor{darkgreen}{rgb}{0.0,0.5,0.0}
\definecolor{brown}{rgb}{0.0,0.0,0.0}

\newcommand{\blue}{\color{darkblue}}

\newcommand{\bta}{{\blue \eta}}

\begin{document}

\begin{titlepage}

\begin{flushright}
\end{flushright}

\vspace{0.2cm}
\begin{center}
\Large\bf
{\boldmath Precision direct photon and $W$-boson spectra at high $p_T$ and comparison to LHC data}
\end{center}

\vspace{0.2cm}
\begin{center}
{\sc Thomas Becher$^a$, Christian Lorentzen$^a$ and Matthew D. Schwartz$^{b}$}\\
\vspace{0.4cm}
{\sl ${}^a$\,Albert Einstein Center for Fundamental Physics\\
Institut f\"ur Theoretische Physik, Universit\"at Bern\\
Sidlerstrasse 5, CH--3012 Bern, Switzerland\\[0.4cm] 
$^b$\,Department of Physics\\
Harvard University\\
Cambridge, MA 02138, U.S.A.}
\end{center}

\vspace{0.2cm}
\begin{abstract}
\vspace{0.2cm}
The differential $p_T$ spectrum for vector boson production is computed 
at next-to-leading fixed order and including the resummation of threshold logarithms at next-to-next-to-leading logarithmic accuracy. A comparison is made to {\sc atlas} data on direct photon and $W$ production at high transverse momentum $p_T$, finding excellent agreement. The resummation is achieved by
factorizing contributions associated with different scales using Soft-Collinear Effective Theory. Each part is then calculated perturbatively and the individual contributions are combined
using renormalization group methods. A key advantage of the effective theory framework is that it indicates a set of natural scale choices, in contrast to the fixed-order calculation.
Resummation of logarithms of ratios of these scales
leads to better agreement with data and reduced 
theoretical uncertainties.
\noindent 
\end{abstract}
\vfil

\end{titlepage}

\section{Introduction}
Cross sections for vector boson production are relatively clean observables at hadron colliders. The differential $p_T$ spectra for photons, $W$ bosons and $Z$ bosons, in particular, provide excellent benchmarks to test the standard model as well as to measure parton-distribution functions (PDFs). An important application of such measurements is to compare and validate different precision calculations, performed at fixed order or including resummation. Vector boson production, therefore, gives us a rare handle to gauge the importance of higher order perturbative effects and power corrections. In this paper, we perform such a comparison on the high $p_T$ photon~\cite{Aad:2011tw} and $W$ boson spectra~\cite{Aad:2011fp} measured by the {\sc atlas} collaboration at the LHC, using around 35 pb${}^{-1}$ data.

Direct (or prompt) photon production is the production of a hard photon in association with a jet. The cleanest direct photon observable is the inclusive photon $p_T$ spectrum, which can be measured independently of any jet definition. At low $p_T$,
there is a large background from $\pi^0$ and other hadronic decays, which are often corrected for
by demanding that the photon be isolated. In the {\sc atlas} study~\cite{Aad:2011tw}, the isolation criteria was that there should be
no radiation with less than 4 GeV of energy in a cone of radius $R=0.4$ around the photon. An advantage
of studying the direct photon spectrum at high $p_T$ is that there is little background of a hard photon
coming from background processes and isolation becomes unnecessary. Formally, the backgrounds
provide only power corrections in this region.

$W$ and $Z$ production have smaller cross sections than photons, especially after paying the cost of a branching ratio to leptons, but do not require isolation. The $W$ spectrum is particularly challenging to measure since it requires an understanding of the missing energy, which unlike the lepton $p_T$, requires mastery of systematic effects over the entire detector. 

The photon, $W$ and $Z$ production rates have been known at the next-to-leading order (NLO) for some time~\cite{Aurenche:1983ws,Aurenche:1987fs,Gordon:1993qc,Ellis:1981hk,Arnold:1988dp,Gonsalves:1989ar}. In this paper, we take leading order (LO) to refer to the leading order in which the vector boson $V$ has non-zero $p_T$. So this is a tree-level $2\to 2$ scattering process. NLO is one order beyond this, which includes 1-loop corrections to the $2 \to 2$ processes as well as $2\to3$ real emission graphs. While the inclusive $V$ production rates are known at NNLO, the differential $p_T$ spectra are only known at NLO. These corrections are implemented in Monte Carlo integration programs to provide the NLO distributions, such as {\sc qt}~\cite{qt}, {\sc mcfm}~\cite{mcfm},  {\sc fewz}~\cite{Melnikov:2006kv,Gavin:2010az}, and {\sc dynnlo}~\cite{Catani:2009sm}.

Beyond NLO, the theoretical calculation of the vector boson spectrum is extremely challenging, and the NNLO result is not yet known. In the absence of this result, one can improve on NLO by adding in partial results at higher orders. In some cases, such as at low $p_T$, this is absolutely critical. The fixed-order calculation diverges at small $p_T$ so one needs to resum logarithms of the form $\ln(p_T/M_V)$ to get even qualitative agreement with data. The resummation at low $p_T$ has been performed at the next-to-next-to-leading logarithmic level (NNLL)~\cite{Balazs:1997xd,Bozzi:2010xn,Becher:2010tm}. 

Also at very high $p_T$, large
logarithms arise, now of the form $\ln(1-p_T/\ptmax)$, where $\ptmax$ is the maximum kinematically possible transverse momentum for the vector boson at a given rapidity. For the photon $\ptmax = \frac{\ecm}{2 \cosh y}$, where $\ecm$ is the machine center-of-mass energy (7 TeV for the 2010-2011 LHC run) and $y$ is the photon's rapidity.
The approach to improving on the fixed-order NLO calculation at high $p_T$ discussed in
~\cite{Laenen:1998qw,Becher:2009th,Becher:2011fc} was to expand around the limit $p_T = \ptmax$.
This is the {\it machine threshold limit}. 
When $p_T = \ptmax$, there is only phase space for the vector boson to be recoiling
against a single parton, which also has $p_T = \ptmax$. If the boson has slightly less $p_T$, then
the recoiling hadronic radiation must be jet-like, with the partons in the jet being either collinear
or soft. Thus it is natural to describe the region near the machine threshold using Soft-Collinear Effective Theory (SCET)~\cite{Bauer:2000yr,Bauer:2001yt,Beneke:2002ph}.

Using traditional methods, the threshold resummation for $W/Z$ production at large $p_T$ was performed at NLL accuracy in~\cite{Kidonakis:1999ur,Kidonakis:2003xm,Gonsalves:2005ng}. Using SCET, the accuracy was increased to NNLL in~\cite{Becher:2009th,Becher:2011fc}. The effective theory approach simplifies the computations, and having operator definitions of the various ingredients of the factorization theorem lets us recycle known results, such as the 2-loop jet function, computed for other applications. This greatly reduces the amount of new analytical results needed. Nevertheless, also in the traditional formalism, the results were recently extended to NNLL accuracy~\cite{Kidonakis:2012su,Kidonakis:2011hm} (although only the NNLO fixed-order expansion of the resummed result was computed in these papers).

In practice, the threshold logarithms are important well away from the machine threshold because of the rapid fall off of the PDFs towards larger values of the momentum fraction $x$ which ensures that most of the cross section comes from a region near the partonic threshold~\cite{Appell:1988ie,Catani:1998tm}. To what extent this dynamical enhancement of the threshold is effective was analyzed in detail in~\cite{Becher:2007ty}. There is a simple phenomenological argument why it should hold: in events with a 300 GeV gauge boson, there is almost always a jet with $p_T\sim$ 300 GeV recoiling against it. That this jet is highly collimated and nearly massless indicates that the phase space region generating the large logarithms relevant for the vector boson $p_T$ spectrum is important. Indeed, in cases such as inclusive Drell-Yan and Higgs production, where the NNLO corrections are known, it is found that 80\,--\,90\% of the perturbative corrections to the cross section arise from the threshold terms, even in cases such as Higgs production, where the fall-off of the PDFs is not very strong. We thus expect our resummed results to provide a good approximation to the full NNLO result.

\section{Effective Field Theory approach}
The effective field theory allows us to obtain logarithmic contributions to the vector boson $p_T$ spectrum which supplement the exact NLO distribution, computed in full QCD. These logarithmic terms arise from the threshold region, where the vector boson has the kinematically maximal transverse momentum. In this region, the jet recoiling against the vector boson is nearly massless. The formal derivation of these threshold terms is performed in the machine threshold limit, where $x\to 1$ for both PDFs. However, once the logarithms are extracted they can be used as additional information about the cross section in the kinematic region where $x$ has more reasonable values. That is, the same threshold logarithms are present for any $x$, since they come from a perturbative calculation in QCD which factorizes from the non-perturbative PDFs. The only difference is that away from the machine threshold, we are no longer guaranteed that the threshold terms dominate the hadronic cross section parametrically. In practice they still give rise to the bulk of the cross section thanks to the dynamical threshold enhancement discussed above. Physically, the threshold logarithms are associated with collinear radiation in the recoiling jet or soft radiation coming from the jet or the incoming partons. A method to sum these logarithms to all orders in perturbation theory using the renormalization group (RG) in SCET was developed in~\cite{Becher:2006nr,Becher:2006mr}. Its application to direct photon production was discussed in detail in~\cite{Becher:2009th}. Here we will only briefly summarize the method.

A simple variable to use for the expansion near the machine threshold is $M_X$, the mass of everything-but-$V$, where $V$ refers to the vector boson ($\gamma,W$ or $Z$). In terms of the proton momenta $P_1^\mu$ and $P_2^\mu$ and the vector boson momentum $q^\mu$,
\begin{equation}
M_X^2 = (P_1+P_2 +q)^2\,.
\end{equation} 
Since $P_1$ and $P_2$ are fixed, $M_X$ is determined completely by the momentum of the vector boson. As its transverse
momentum approaches its maximum allowed value at fixed rapidity, $M_X \to 0$.
To understand the relevant degrees of freedom, it is helpful also to consider the partonic version of
$M_X$, called $m_X$. This is defined as
\begin{equation}
m_X^2 = (p_1+p_2 +q)^2\,,
\end{equation} 
where $p_1^\mu = x_1 P_1^\mu$ and $p_2^\mu = x_2 P_2^\mu$ are the momenta of the partons coming out of the protons which participate in the hard interaction. Taking  $m_X\to 0$ is called the {\it partonic threshold limit}. Obviously, $M_X \to 0$ implies $m_X \to 0$. The partonic $m_X$ is like $M_X$ without including the beam remnants. Away from the machine threshold, the beam remnants make $M_X$ large while $m_X$ can remain small. Thus the logarithms we actually expect to be important in affecting the vector boson $p_T$ spectrum beyond NLO can be deduced by considering the theoretically simpler but less physical partonic threshold limit.

Near the partonic threshold, the vector boson must be recoiling against a jet and there is only phase space for the jet to be nearly massless.  So then we can write
\begin{equation}
m_X^2 = (p_J + k_S)^2 \approx   p_J^2 +2E_J k\,, 
\end{equation} 
where $p_J^\mu$ and $k_S^\mu$ are the collinear and soft momenta in the jet,
$E_J$ is the jet energy and $k=p_J\cdot k_S/E_J$. It is because of this decomposition that the logarithmic terms we will extract come from either collinear effects ($p_J^2 \to 0$) or soft effects $k\to 0$.  SCET implements the structure of the soft and collinear emissions on the Lagrangian level, using different fields to describe the soft and collinear partons. Via a field redefinition, the two sectors can be decoupled, after which the soft emissions are obtained from soft Wilson lines running along the directions of large momentum.

\begin{figure}[t!]
\begin{center}
\psfrag{p1}[B]{$p_1$}\psfrag{p2}[t]{$p_2$}\psfrag{pJ}[t]{$p_J$}
\psfrag{q}[B]{$q$}
\begin{tabular}{ccc}
\multirow{4}{*}{ \includegraphics[width=0.5\textwidth]{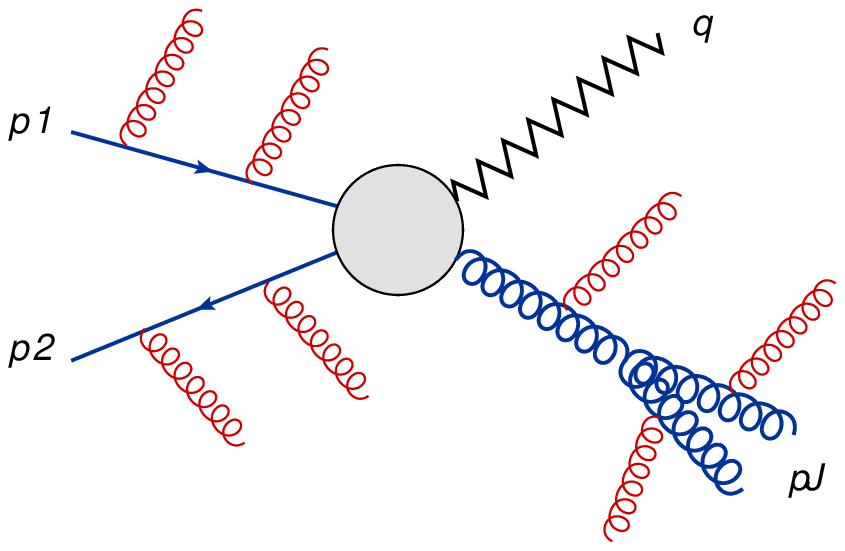}} && \\[-1cm]
&\phantom{abcdef}& \includegraphics[height=0.077\textwidth]{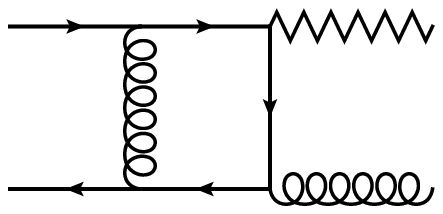}\\[0.3cm]
&&\includegraphics[height=0.11\textwidth]{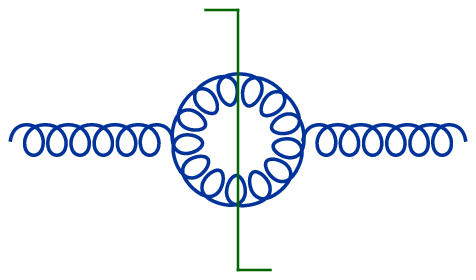}\\[0.3cm]
&& \includegraphics[height=0.11\textwidth]{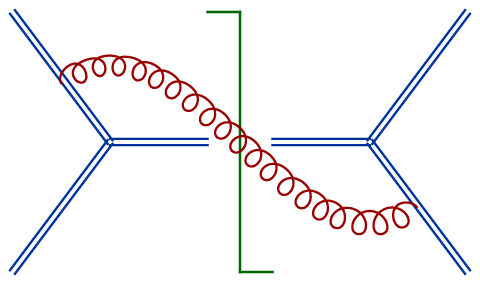}
\end{tabular}
\end{center}
\caption{Left: Factorization of the scattering amplitude near the partonic threshold. Right: Examples of NLO corrections to the hard, jet and soft function (from top to bottom). The thick blue lines denote  partons collinear to the directions of the jet or the incoming hadrons. Soft emissions are pictured by thin red gluon lines. 
\label{fig:factheorem}}
\end{figure}

The result from SCET is that the partonic cross section in the threshold region for any particular
channel has the form
\begin{equation} \label{fform}
\hat{s} \frac{\rd\hat{\sigma} }{\rd\hat{u}\, \rd\hat{t}} = \hat{\sigma}^{(0)}( \hat{u},\hat{t})\, 
H (\hat{u},\hat{t},M_V,\mu)
 \int\! \rd k\, J (m_X^2-2 E_J k) S(k,\mu)\, ,
\end{equation} 
where the partonic Mandelstam variables are $\hat{s} = (p_1+p_2)^2$, $\hat{t} = (p_1-q)^2$ and $\hat{u} = (p_2-q)^2$, with $q$ the vector boson momentum,
with $q^2=M_V^2$.
 We have factored out the Born level cross section $ \hat{\sigma}^{(0)}(\hat{u},\hat{t})$. The hadronic cross section is obtained after convoluting with PDFs and summing
over all partonic channels (see Sec. \ref{sec:intvars} below).

The factorization theorem in Eq.~\eqref{fform} is depicted in Figure~\ref{fig:factheorem}. The hard function $H$ contains the virtual corrections to the underlying hard-scattering process.  There are two channels relevant for vector boson production, the Compton ($q g \to V q$) and annihilation ($q \bar{q} \to V g$) channels, and the corresponding hard functions are related by crossing symmetry. A sample NLO contribution to the hard function in the annihilation channel is the top one-loop diagram on the right-hand side of Figure~\ref{fig:factheorem}. For the photon case, the one-loop hard function was given in~\cite{Becher:2009th}, and in~\cite{Becher:2011fc} it was outlined how the hard function can be obtained for $M_V \ne 0$. For completeness, we list the one-loop result for both the Compton and annihilation channel in the Appendix.  The jet function $J$ encodes the collinear emissions inside the final state jet, while collinear emissions along the initial state partons are absorbed into the PDFs. The jet function is obtained from the imaginary part of the two-point function of collinear fields (see the middle Feynman diagram on the right in Figure~\ref{fig:factheorem}). The two-loop results for the  inclusive quark and gluon jet functions relevant here were obtained in~\cite{Becher:2006qw} and~\cite{Becher:2010pd}. The last Feynman diagram in the figure shows a NLO correction to the soft function, which describes the soft emissions from the energetic partons in both the initial and final state, which are encoded in Wilson lines along the corresponding directions. The corresponding soft function was recently computed to two loops in~\cite{Becher:2012za}.

In the remainder of this section, we give the resummed result for the cross section and discuss its numerical implementation. We first set up the integration over the parton momentum fractions in a form suited for threshold resummation and then give the resummed result, as well as the matching to fixed-order perturbation theory. Finally, we discuss how subtractions can be used to improve the convergence of the numerical integrations.

\subsection{Integration variables\label{sec:intvars}}
The perturbative calculation, whether NLO or including resummation, produces partonic cross sections. The observable boson $p_T$ spectrum is then obtained after convoluting with PDFs, 
\begin{equation}
\frac{\rd^2 \sigma}{\rd y\, \rd p_T} 
=  \sum_{ab}
\int_0^1 \rd x_1 \int_0^1 \rd x_2
 f_{a/N_1} (\x_1, \mu)  f_{b/N_2} (\x_2, \mu) 
\frac{\rd^2 \hat \sigma_{ab}}{\rd y \,\rd p_T} \,,
\end{equation}
where the sum is over all partonic channels, $a,b \in \{q,\bar{q}, g\}$. The partonic cross section can also be written as
\begin{equation}
\frac{\rd^2 \hat \sigma_{ab}}{\rd y \rd p_T^2} =\hat s\,\frac{\rd^2 \hat \sigma_{ab}}{\rd \hat t \rd \hat u}\,.
\end{equation}
At NLO, in a given channel, it has the general form
\begin{equation} \label{nloform}
\hat{s}\frac{\rd^2 \hat \sigma}{\rd \hat t\, \rd \hat u} = \hat{\sigma}^{(0)}  \left\{ \delta(\Mxt) + \alpha_s(\mu)\left[ \delta(\Mxt)\, h^{(1)} + \left[ \frac{1}{\Mxt} \right]_\star^{[\mu]}  h^{(2)}
+ \left[\frac{\ln\frac{\Mxt}{\mu^2}}{\Mxt} \right]_\star^{[\mu]} h^{(3)} + h^{(4)}  \right]\right\}
\, ,
\end{equation}
where $\hat{\sigma}^{(0)} $ and the coefficients $h_i$ are functions of the two variables $\hat t$ and $\hat u$. Because of the relation  $\hat s+ \hat t+ \hat u = \Mxt +M_V^2$  the $\delta$-function parts, effectively, only depend on a single variable. The $\star$-distributions are generalizations of the usual $+$-distributions to dimensionful variables~\cite{De Fazio:1999sv}. A N$^n$LO computation
would give distributions with logarithms up to $\ln^{2n-1}(\Mx/\mu)$ in the numerator. Resummation allows one to predict these singular terms at higher orders, but not the regular parts, such as $h^{(4)}$.

The leading-order cross sections for the production of a photon are
\begin{align}\label{born}
\hat\sigma^{(0)} _{\qq} &= \frac{2\, C_F\,\pi\, \alpha_{
\rm e.m.} \alpha_s(\mu)}{N_c \hat{s}}  \, e_q^2\,  T_0(\hat{u},\hat{t}),
 & \hat\sigma^{(0)} _{qg}&= -\frac{\pi\, \alpha_{
\rm e.m.} \alpha_s(\mu)\,}{N_c \hat{s}}\, e_q^2\, T_0(\hat{s},\hat{t}) \, ,
\end{align}
where $e_q$ is the charge of the quark and 
\begin{equation}
 T_0(u,t) = \frac{u}{t}+\frac{t}{u}+\frac{2 M_V^2\, (M_V^2-t-u)}{t u}\,.
\end{equation}
For the photon $M_V^2=q^2=0$, but we need the same expression also for $Z$ and $W$ bosons. The amplitude $\hat\sigma^{(0)}_{gq}$ is obtained by replacing $T_0(\hat s,\hat t) \to T_0(\hat s,\hat u) $ in the expression for  $\hat\sigma^{(0)} _{qg}$. To obtain the amplitude for $Z$ production, one replaces the quark charge in (\ref{born}) by
\begin{equation}
  e_q^2 \to\frac{|g_L^q|^2+|g_R^q|^2}{2}
   = \frac{\big( 1 - 2|e_q|\sin^2\theta_W \big)^2 + 4 e_q^2\sin^4\theta_W}%
                  {8\sin^2\theta_W\cos^2\theta_W} \,,
\end{equation}  
where $\theta_W$ is the weak mixing angle. Since the $W$ bosons have flavor-changing couplings, the sum over flavors must be replaced by a double sum over individual quark and antiquark flavors, $q$ and $q'$. Only left-handed currents appear in this case. The relevant coupling for a $W^-$ boson produced in the annihilation of an anti-up and a down quark is
\begin{equation}
  \frac{|g_L^{q'q}|^2}{2}
   = \frac{|V_{q'q}|^2}{4\sin^2\theta_W} \,,
\end{equation}
where $V_{q'q}$ are elements of the quark mixing matrix.

Because of the singular behavior of the partonic cross section at threshold, it is advantageous to introduce $\Mxt$ as an integration variable.  Following~\cite{Ellis:1981hk}, we perform the integrations in the form
\begin{equation}
\frac{\rd^2 \sigma}{\rd y \rd p_T} 
=   \sum_{ab}
\int_{x_{\rm min}}^1 \frac{\rd x_1 }{x_1 s+u- M_V^2 }\int_0^{m_{\rm max}^2} \rd m_X^2
  f_{a/N_1} (\x_1, \mu)  \, f_{b/N_2} (\x_2, \mu) \frac{\rd^2 \hat\sigma_{ab}}{\rd y \rd p_T} \, , \label{newvar}
\end{equation}
with
\begin{align}
 u & =(P_2-q)^2 = M_V^2-\sqrt{s}\sqrt{M_V^2+p_T^2}e^y \, , \nonumber\\
 m_{\rm max}^2 &=u + x_1 (M_X^2- u) \, , \\
 x_{\rm min} &=\frac{-u}{M_X^2-u}\,.\nonumber
\end{align}
When performing the resummation, one performs an expansion of the cross section around the partonic threshold $\Mx = 0$. With the choice of variables adopted in Eq.~(\ref{newvar}), the expansion is performed at fixed $\x_1$ or, equivalently, at fixed $\hat t = (p_1-q)^2$. This is problematic, since the expansion then induces unphysical rapidity asymmetries. In order to avoid this and obtain a symmetric form, we integrate twice: first with Eq.~(\ref{newvar}) in the variables $x_1$ and $\Mx$ and then with the $ u \leftrightarrow  t$ crossed version of  Eq.~(\ref{newvar}) in the variables $x_2$ and $\Mx$. By taking the average of these two results, we obtain a symmetric form of the expansion around the threshold. In the case of direct photon production, a more convenient choice of integration variables is  
\begin{equation}
v = 1 + \frac{\hat{t}}{\hat{s}}
\, , 
\hspace{1em}
w = -  \frac{\hat{u}}{\hat{s} + \hat{t}}\,.
\end{equation}
In this case, the partonic threshold is at $w=1$. Using these variables significantly improves the numerical integration. The resummed photon cross section in $v$ and $w$ was given in~\cite{Becher:2009th}.

\subsection{Resummation and matching to fixed order}

\begin{figure}[t!]
\begin{center}
\psfrag{m}[t]{}
\psfrag{h}{$\mu_h$}
\psfrag{j}{$\mu_j$}
\psfrag{s}{$\mu_s$}
\psfrag{f}{$\muf
$}
\psfrag{H}[]{$H_{I}(\hat{u},\hat{t})$}
\psfrag{J}[]{$J_{I}(m_X^2)$}
\psfrag{S}[]{$S_{I}(k)$}
\psfrag{F}[l]{$f_1(x_1)f_2(x_2)$}

\includegraphics[width=0.6\hsize]{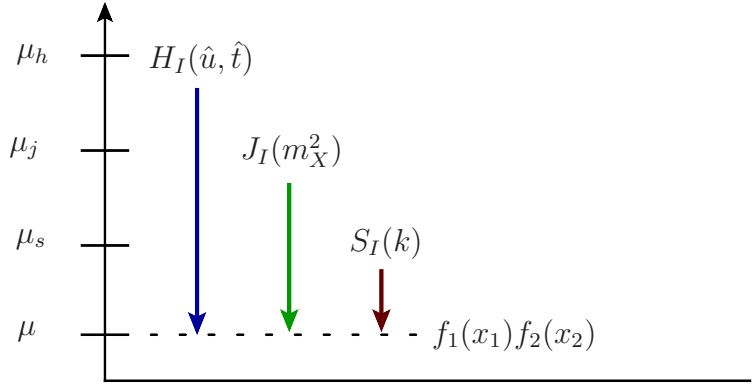}
\end{center}
\vspace{-0.5cm}
\caption{Resummation by RG evolution.\label{running}}
\end{figure}

The resummed result for the  cross section is obtained by solving the RG equations for the hard, jet and soft functions. Each function is then evaluated at its natural scale, where it does not suffer from large logarithmic corrections, and evolved to a common scale $\mu$, which we identify with the factorization scale, see Figure~\ref{running}. The solution of the RG equation for the hard function in the annihilation channel is
\begin{equation}
\hat\sigma^{(0)}_{\qq}(\hat u,\hat t,\mu)\, H_{\qq}(\hat u,\hat t,\mu) = U_{H_{\qq}}(\mu_h,\mu) \, \hat\sigma_{\qq}^{(0)}(\hat u,\hat t,\mu_h) H_{\qq}(\hat u,\hat t,\mu_h) \,.
\end{equation}
The matching scale $\mu_h$ is chosen to be of the order of $p_T$ to avoid large logarithms. The NLO result for $H_{\qq}(\hat u,\hat t,\mu_h)$ was given in~\cite{Becher:2011fc}.
We have included the Born cross section which depends on the scale via the coupling constant $\alpha_s(\mu)$. The evolution factor $U_{H_{\qq}}(\mu_h,\mu)$ for the above combination takes the form
\begin{equation}
\ln U_{H_{\qq}}(\mu_h,\mu) = 2 \left( C_F + \frac{C_A}{2} \right) \left [ 2 S(\mu_h,\mu) - A_{\rm cusp}(\mu_h,\mu) \ln\frac{\hat{s}}{\mu_h^2}\right] - 2 A_{{H_{\qq}}}(\mu_h,\mu)\,,   \,
\end{equation}
with
\begin{align}
S (\nu, \mu) &= - \int_{\alpha_s (\nu)}^{\alpha_s (\mu)} d \alpha
\frac{\gamma_{\mathrm{cusp}} (\alpha)}{\beta (\alpha)} \int_{\alpha_s
(\nu)}^{\alpha} \frac{\rd \alpha'}{\beta (\alpha')}\,, &
A_{\rm cusp} (\nu, \mu) &= - \int_{\alpha_s (\nu)}^{\alpha_s (\mu)} d \alpha
\frac{\gamma_{\mathrm{cusp}} (\alpha)}{\beta (\alpha)}\,.
\end{align}
Explicit expressions for these functions in RG-improved perturbation theory can be found in~\cite{Becher:2006mr}. The function $A_{H_{\qq}}$ is the same as $A_{\rm cusp} (\nu, \mu)$ with \begin{equation}
\gamma_{H_{\qq}} = 2 \gamma_q+ \gamma_g - \frac{C_A}{2} \ln \frac{\hat{s}^2}{\hat{t} \hat{u}}\, \gamma_{\rm cusp}
\end{equation}
replacing
$\gamma_{\mathrm{cusp}}$.  
The quark and gluon anomalous dimensions $\gamma_q$ and $\gamma_g$ are given to three-loop order in~\cite{Becher:2009qa}. The evolution factor $U_{H_{qg}}$ can be obtained from the above results using the crossing relation  $\hat{s} \leftrightarrow -\hat{u}$ at fixed $\hat t$ , and $U_{H_{gq}}$ follows from $U_{H_{qg}}$ using $\hat{t} \leftrightarrow \hat{u}$. The resummed results for the jet and soft functions can be obtained by solving their RG equations in Laplace space ~\cite{Becher:2006nr}. For the gluon jet function, for example, the result takes the form
\begin{align} \label{eq:RGjet}
J_g (p^2, \mu) &= U_{J_g}(\mu_j,\mu) \,\widetilde{j}_g (\partial_{\blue \eta_{j_g}}) \frac{1}{p^2} \left(
\frac{p^2}{\mu_j^2} \right)^{\blue \eta_{j_g}} \frac{e^{- \gamma_E {\blue \eta_{j_g}}}}{\Gamma ({\blue \eta_{j_g}})}\,, \nonumber
\end{align}
where $\widetilde{j}_g$ is the Laplace transform of the momentum-space jet function and
\begin{align}
U_{J_g}(\mu_j,\mu) &=  \exp [- 4 C_A S (\mu_j, \mu) + 2 A_{J_g} (\mu_j, \mu)] \,, \\
{\blue \eta_{j_g}} &= 2 C_A A_{\rm cusp} (\mu_j, \mu)  \non \, .
\end{align}
The corresponding results for the quark jet function and the soft functions in the different channels are all listed in~\cite{Becher:2009th}, together with the necessary anomalous dimensions. Inserting the resummed expressions into (\ref{fform}), one  obtains
\begin{align}
\frac{\rd^2 \hat\sigma_\qq}{\rd y \rd p_T^2} &=  \sigma^{(0)}_{\qq}(\hat u,\hat t,\mu_h)\,H_{\qq}(\hat u,\hat t,\mu_h) \, U_{H_{\qq}}(\mu_h,\muf
)\, U_{J_g}(\mu_j,\muf
)\, U_{S_{\qq}}(\mu_s,\muf
) \nonumber \\
&
\phantom{==} \times  \widetilde{j}_g (\partial_{\blue \eta_{j}}, \mu_j)\, \frac{1}{\Mxt} \left( \frac{\Mxt}{\mu_j^2} \right)^{\blue \eta_j} \, \widetilde{s}_{\qq} ( \partial_{\blue \eta_{s}},
\mu_s)\,  \left( \frac{\Mxt}{ E_h \mu_s} \right)^{\blue \eta_s}\,
\frac{e^{- \gamma_E \left({\blue \eta_{j}}+ {\blue \eta_{s}}\right) }}{
\Gamma \left({\blue \eta_{j}}+ {\blue \eta_{s}}\right) } \, , \nonumber\\
&=  \sigma^{(0)}_{\qq}(\hat u,\hat t,\mu_h)\,H_{\qq}(\hat u,\hat t,\mu_h) \, U_{H_{\qq}}(\mu_h,\muf
)\, U_{J_g}(\mu_j,\muf
)\, U_{S_{\qq}}(\mu_s,\muf
) \nonumber \\
&
\phantom{==} \times
 \left( \frac{\mu_j^2}{E_h\mu_s} \right)^{\blue \eta_s}
 \widetilde{j}_g (\partial_{\blue \eta_{\qq}}, \mu_j)\,  \widetilde{s}_{\qq} \Big( \partial_{\blue \eta_{\qq}}+\ln\frac{\mu_j^2}{E_h \mu_s},\mu_s\Big)\,  \frac{1}{\Mxt} \left( \frac{\Mxt}{\mu_j^2} \right)^{\blue \eta_\qq}
\frac{e^{- \gamma_E {\blue \eta_{\qq}}}}{\Gamma \left({\blue \eta_{\qq}}\right) } \, ,\label{resultsigma}
\end{align}
where $E_h = \sqrt{{\hat{t}\hat{u}}/{\hat{s}}} = p_T$ and
\begin{equation}
{\blue \eta_{\qq}} ={\blue \eta_{j_g}} +{\blue \eta_{s_{\qq}}}  
= 2 C_A A_{\rm cusp} (\mu_j, \mu) + (4 C_F - 2 C_A)
A_{\rm cusp} (\mu_s, \mu)\,. \non
\end{equation}
To arrive at this expression, the convolution integral in (\ref{fform}) was explicitly carried out, which is possible because of the simple form of the RG evolved soft and jet functions. 

Using the general expression Eq.~\eqref{resultsigma} for the resummed cross section, we can now explicitly compute the resummed distribution. We include almost all of the ingredients for N${}^3$LL accuracy. For N${}^3$LL, one needs the cusp anomalous dimension to four-loop order and the regular anomalous dimensions to three loops, together with the two-loop results for the hard, jet and soft functions. For the anomalous dimensions, the only missing ingredient is the unknown four-loop cusp anomalous dimension, which we estimate using the standard Pad\'e approximation as $\Gamma_4 = \Gamma_3/ (\Gamma_2)^{2}$. The second ingredient, which we do not include  are the non-logarithmic pieces of the jet, soft, and hard functions. The full two-loop jet functions are known~\cite{Becher:2006qw,Becher:2010pd} and also the two-loop soft function has now been computed \cite{Becher:2012za}. The non-logarithmic piece of the two-loop hard function can be extracted from the results of~\cite{Garland:2002ak,Gehrmann:2011ab}, and we plan to include the full two-loop matching in the future. To indicate that we only have a partial result, we denote our highest order by N${}^3$LL$_{\rm p}$.

In order to include all the available perturbative information, we match to the NLO fixed-order result, which is the highest order available. To perform the matching, we use~\cite{Becher:2007ty}
\begin{equation}\label{match}
\left(\frac{\rd^2\sigma}{\rd y \rd p_T^2}\right)^{\mathrm{N^3LL_{p} + NLO}} 
= \left(\frac{\rd^2\sigma}{\rd y \rd p_T^2}\right)^{\mathrm{N^3LL_{p}}}   +  \left(\frac{\rd^2\sigma}{\rd y \rd p_T^2}\right)^{\mathrm{NLO}} - 
\left(\frac{\rd^2\sigma}{\rd y \rd p_T^2}\right)^{\mathrm{NNLL}}_{\mu_h=\mu_j=\mu_s=\muf
}\,.
\end{equation}
The subscript on the last  term indicates that all
scales must be set equal to the relevant value of $\muf
$. Setting these scales equal switches off the resummation. The NNLL expression includes the one-loop corrections to the hard, jet and soft functions. Once it is evaluated with all scales equal, it reduces to the singular threshold terms of the NLO result, which must be subtracted since they are already included in the resummed result.

\subsection{Subtractions}
To compare with data, one needs to perform 4-dimensional integrals, over $x_1$, $\Mx$, $y$ and $p_T$. These numerical integrals
are computationally expensive and additionally challenging because of the singular nature of the partonic cross sections. After resummation the partonic cross sections are no longer distribution valued at the partonic threshold, but behave as
\begin{equation}
 \frac{1}{\Mxt} \left( \frac{\Mxt}{\mu_j^2} \right)^{\blue \eta}
\end{equation}
near the threshold, see (\ref{resultsigma}). For the natural hierarchy $\mu_j\geq \mu_s \geq \mu $ the quantity $\bta$ is larger than zero and the integral over $\Mx$ converges. To see this, one rewrites ${\blue \eta_{\qq}}$ in the form
\begin{equation}
{\blue \eta_{\qq}} = 2 C_A \,A_{\rm cusp}\, (\mu_j, \mu_s) + 4 C_F \,A_{\rm cusp} (\mu_s, \mu)\, .
\end{equation}
In practice, however, the scale hierarchy is not very large, so that convergence can be quite slow. Furthermore, we will choose a high value of the factorization scale $\mu$, in which case the integral is no longer guaranteed to exist since ${\blue \eta_{\qq}}$ can become negative.

For some threshold variable $m$, the integral we need to evaluate has the form
\begin{equation}
  I (M) = \int_0^M \rd m\, m^{2\bta - 1} f (m)\,.
\end{equation}
To analytically continue the result to negative $\bta$ values, or to improve convergence, it is useful to perform subtractions~\cite{Becher:2007ty,Becher:2006mr}.  A single subtraction would use
\begin{align}
  I (M) &= \int_0^M \rd m \,m^{2\bta - 1} \Big\{ f (0) + \left[ f (m) - f (0)\right] \Big\}
  \\
  &= \int_0^M d m \left\{ \frac{1}{2\bta} M^{2\bta - 1} f (0) + m^{2\bta - 1}
  \left[ f (m) - f (0) \right] \right\}\,,
\end{align}
where the difference $f(m)-f(0)$ makes the integral more convergent. Indeed, assuming $f(m)$ is smooth,
the integral will now converge for $\bta>-1/2$.  A second subtraction would give
\begin{equation}
  I (M) = \int_0^M \rd m \left\{ \frac{1}{2\bta} M^{2\bta - 1} f (0) +
  \frac{1}{2\bta + 1} M^{2\bta} f' (0) + m^{2\bta - 1} \left[ f (m) - f (0) -
  m f' (0) \right] \right\}\,,
\end{equation}
which makes the integral converge for $\bta > -1$. Analogously, one can perform higher-order subtractions to make the integral more and more convergent. In practice, performing too many subtractions slows the code down because the expressions become lengthy and the subtraction itself can become numerically unstable. 
Generally, we have found that using two or three subtractions gives stable results for the N${}^3$LL resummed integrand. Note that higher subtractions involve derivatives of PDFs, which we compute by first interpolating the PDFs.

\section{Scale setting}
The resummed distribution obtained using SCET involves four matching scales: the hard scale $\mu_h$, the jet scale $\mu_j$, the soft scale $\mu_s$ and the the factorization scale $\muf
$. Each ingredient of the factorization formula, thus, can be evaluated at its appropriate scale. This is in contrast to the fixed-order calculation which involves only a single scale, the factorization scale $\muf
$. In addition to the factorization scale, a renormalization scale $\mu_r$ is often introduced by hand in fixed-order computations, by expressing the coupling constant as  $\alpha_s(\mu_r)$. Introducing a second scale $\mu_r$ may be useful as a tool to estimate uncertainties, but there is no physical justification for having $\muf \ne \mu_r$ when
working at fixed order. In contrast, the additional scales in SCET correspond to different physical regions.

If one chooses all the scales equal in the SCET calculation, the resummation is switched off. 
Then only a single RG scale remains.
In this limit, SCET generates all of the terms that are singular in the threshold limit at NLO, as shown in Eq.~\eqref{nloform}. Since we include all logarithmic pieces in the two-loop hard, jet and soft functions, we also obtain all singular terms at NNLO, except for the coefficient of $\delta(m_X^2)$. In general, the singular terms amount to a large fraction of the full perturbative correction. For electroweak boson production at large transverse momentum, they amount to 70--80\% of the NLO correction~\cite{Becher:2011fc}. For inclusive Drell-Yan and Higgs production, also the NNLO correction is known, and it is found that a similarly large fraction of the perturbative corrections arises from the partonic threshold region~\cite{Becher:2007ty,Ahrens:2008nc}.

A perpetual frustration with fixed-order calculations is that they provide no insight  into
how to choose the factorization scale in problems that involve physics at several scales. For example, for $W$ production either $\mu = M_W$ or $\mu=\sqrt{M_W^2 + p_T^2}$ might seem natural. For large $p_T$,  the difference in the prediction between these different parameterizations is larger than the scale variation within any particular parametrization. See for example Figure~\ref{fig:nlobands}, to be discussed more below. If higher order calculations were available, as they might be soon for direct photon or $W$ production, the scale variation and parametrization
dependence would weaken. Unfortunately, however, there are only a handful of observables which have ever been computed beyond NLO. For more complicated processes, such as $W$+4 jet production, NNLO is a distant hope, and in this case there are many possible natural parameterizations. Thus the choice of parametrization for the factorization scale can amount to a significant and difficult-to-estimate source of uncertainty for a fixed-order computation.

\begin{figure}[t!]
\begin{center}\hspace{-0.5cm}
\includegraphics[height=0.3\textwidth]{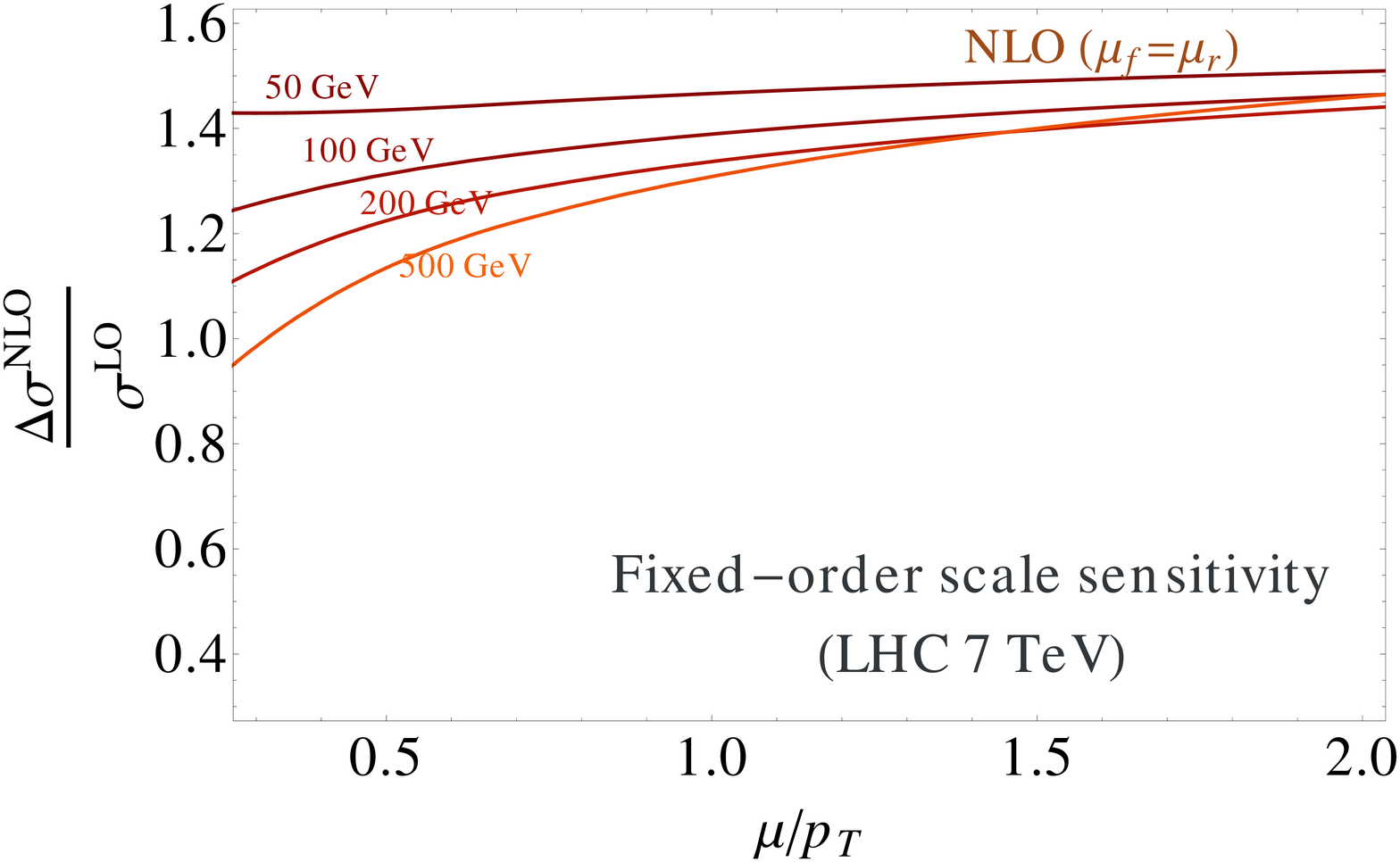}
\hspace{0.2cm}
\includegraphics[height=0.3\textwidth]{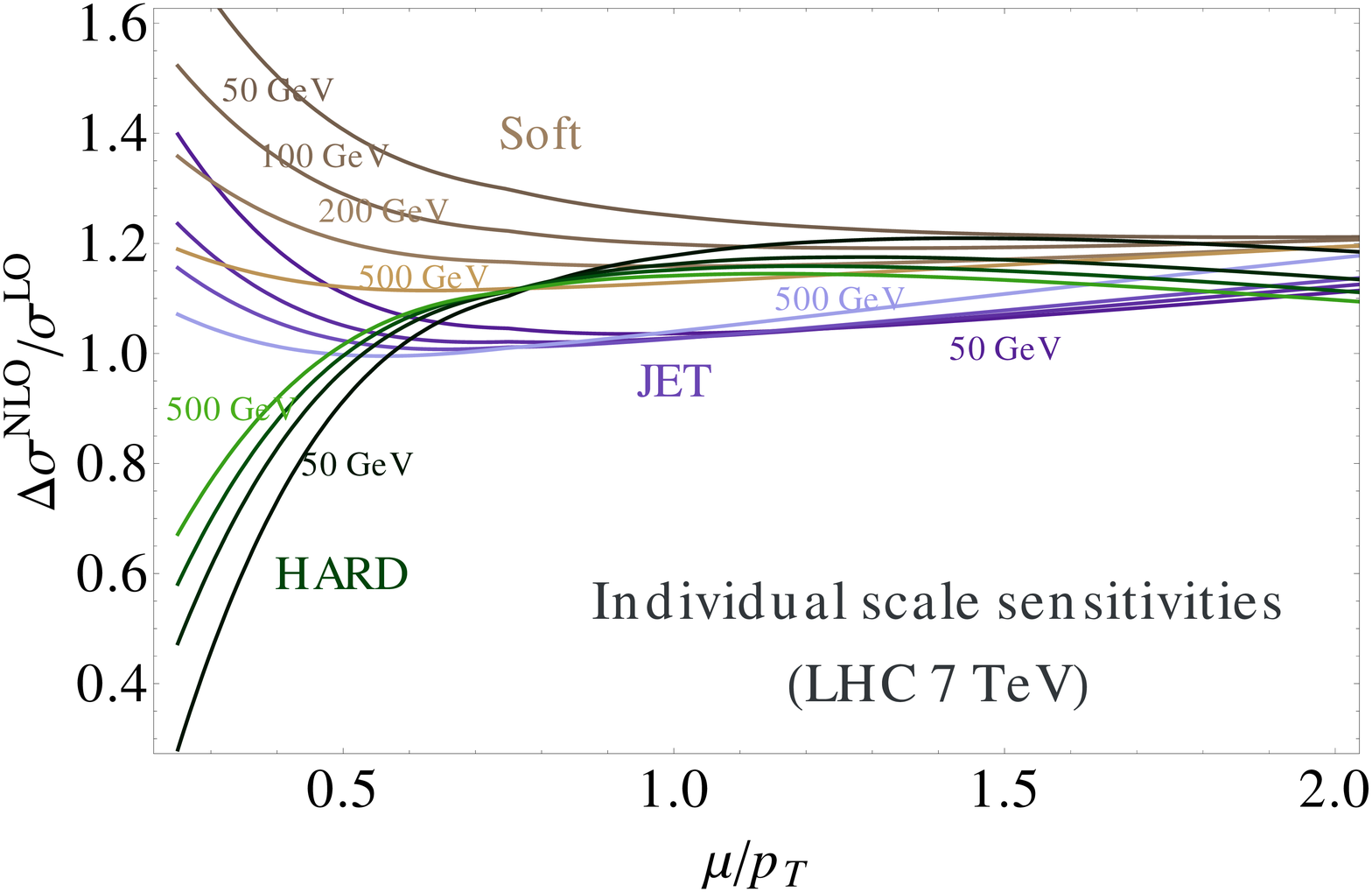} 
\end{center}
\vspace{-0.5cm}
\caption{Scale sensitivities. These plots show the effect of adding part of the fixed-order NLO calculation to the LO calculation. The left panel shows what happens if all the $\mu$-dependent terms at NLO are added together. There is a slow monotonic logarithmic $\mu$ dependence, with no natural extremum. In contrast, when the hard, jet, or soft contributions are added separately, there are natural extrema. These extrema indicate the average value of momenta $\langle p \rangle$ appearing in the logarithms. That there are different extrema for the different components proves that multiple scales are relevant. The plots are for $W^+$ bosons, but the qualitative features are the same for all bosons. \label{fig:scalefig}}
\end{figure}

An extremely satisfying feature of the effective field theory approach is that it {\it does} indicate
what the appropriate parametrization should be. For some observables, such as
$e^+e^-$ event shapes~\cite{Schwartz:2007ib,Becher:2008cf,Chien:2010kc,Feige:2012vc}, the scales are manifest in the resummed distribution. In hadronic collisions, one needs a somewhat more sophisticated procedure since the scales can depend on the functional form of the non-perturbative PDFs. A method for determining these scales without any arbitrary input of what is natural and what is not was proposed in~\cite{Becher:2007ty} and applied to direct photon and $W$ production in~\cite{Becher:2009th,Becher:2011fc}. 
The idea is very simple: one supplements the leading-order calculation with just one part of
the SCET calculation at a time, for example, the hard, jet or soft function evaluated
at  NLO in fixed-order perturbation theory. Doing this, there should be a single scale $\langle p \rangle$ which is the average value of momentum associated with these degrees of freedom, appearing in the large logarithms. After integrating over the PDFs, the perturbative correction will then have the form
\begin{equation}\label{logs}
\frac{\Delta\sigma}{\sigma^{\rm LO}} = \alpha_s(\mu)(c_2 L^2 + c_1 L + c_0)\,,
\end{equation}
with $L=\ln\frac{\mu}{\langle p\rangle}$. If $\mu$ is chosen either much lower or much higher than $\langle p \rangle$, the perturbative corrections will become large. Since we do not have an analytic expression for the distribution, due to the necessity of convoluting
with PDFs, we determine $\langle p \rangle$ numerically by computing the individual corrections to the cross section as a function of $\mu$. The result is shown in the right plot of Figure~\ref{fig:scalefig}. It has the expected form (\ref{logs}) and we see that while the jet and soft scales are concave upwards, the hard curve is concave downward. The extrema of the corresponding curves indicate the scales $\langle p \rangle$ that dominate these contributions after integrating over the PDF. It is then natural to define our default values for $\mu$ as the positions of the extrema.  That there are different extrema for the different components proves that multiple scales are relevant. These scales are conflated in the fixed-order calculation.
The left plot in Figure~\ref{fig:scalefig} shows the fixed-order scale dependence. In this case, there is monotonic $\mu$ dependence, with no natural extremum.

\begin{figure}[t]
\begin{center}
\hspace{-0.3cm}
\includegraphics[width=0.49\textwidth]{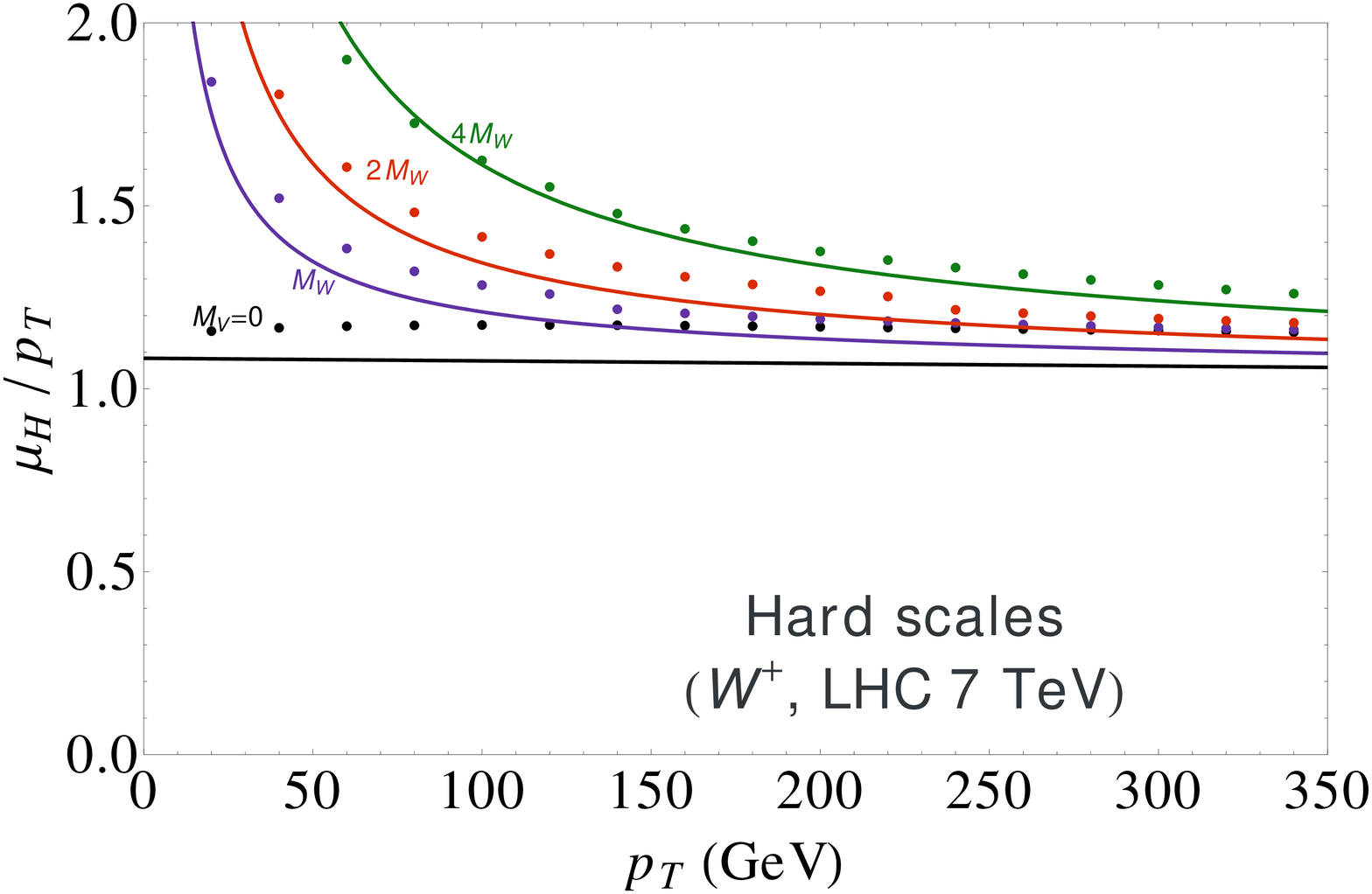}
\includegraphics[width=0.49\textwidth]{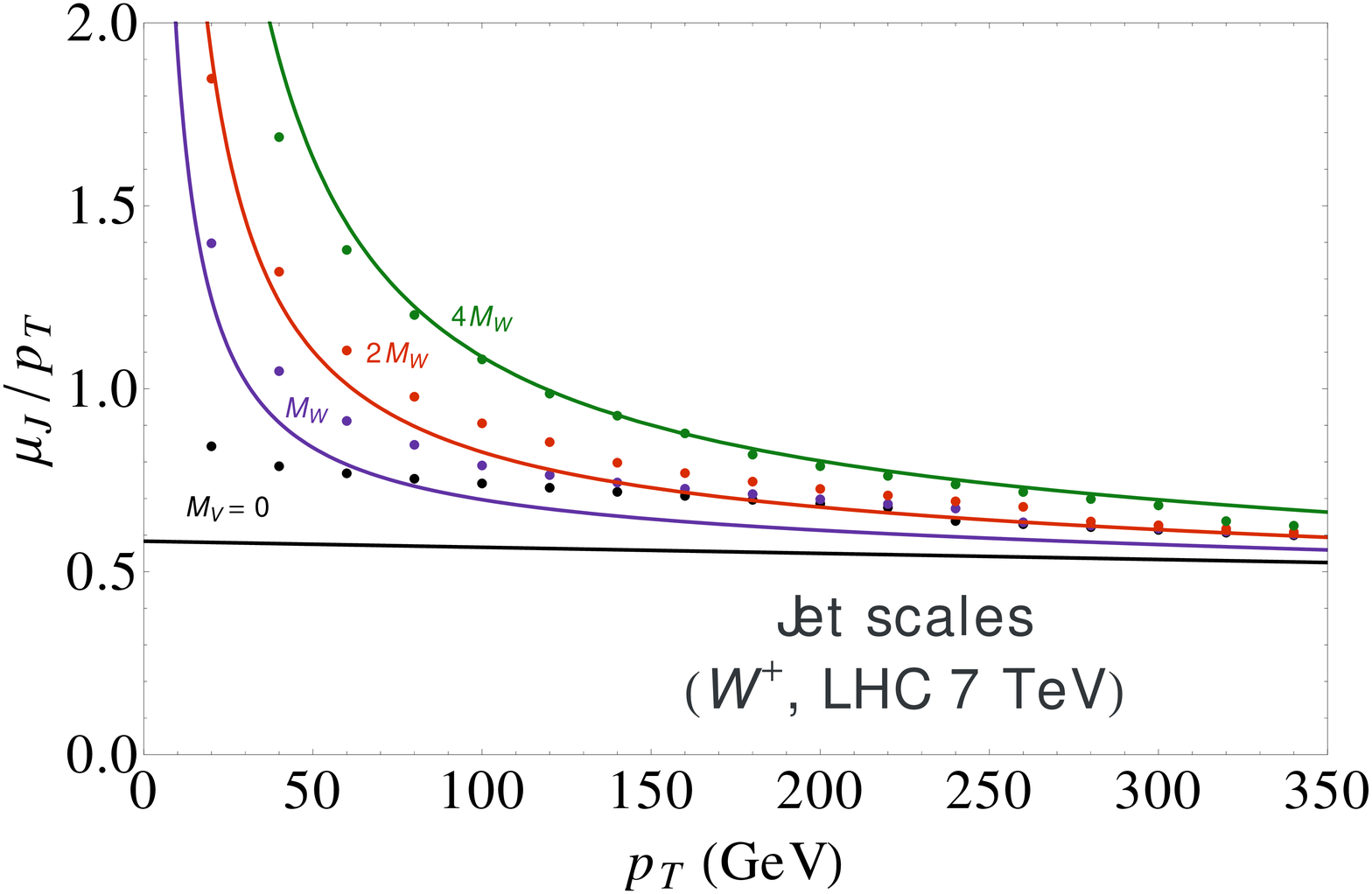}
\end{center}
\vspace{-0.4cm}
\caption{Natural hard and jet scales. These scales are determined, after integrating
over the PDFs, as typical energies appearing in the logarithms when the hard or jet functions at
next-to-leading order only are included. Dots show extrema for various values of boson masses,
and lines show our approximations. An important qualitative point is that the jet scale is naturally lower than the hard scale. This is an output from our numerical procedure, not an input from a formal
analysis.}
\label{fig:scalesnlo}
\end{figure}

To find the scales numerically we extract these extrema from the curves. Using a number of different
machine center-of-mass energies (we tried 2, 7, 14, and 100 TeV), $pp$ and $\bar{p} p$
collisions, and various boson masses, we determine a reasonable approximation to these points is given by
the following functional forms
\pagebreak
\begin{align} \label{scalevalues}
\mu_h &=  \frac{13 p_T+ 2 M_V}{12} -\frac{p_T^2}{\sqrt{s}}\, ,\\ 
\mu_j &= \frac{7 p_T + 2 M_V }{12} \left (1-\frac{2p_T}{\sqrt{s}}\right) \, .
\end{align}
A comparison of this fit to the extrema for the hard and jet scales is shown in Figure~\ref{fig:scalesnlo}.
We have constrained the jet scale to vanish at the endpoint $p_T =\sqrt{s}/2$ since at that point
there is no phase space for collinear emission and the recoiling jet must be massless. 
An alternative and slightly simpler hard scale choice that is quantitatively 
equivalent for LHC energies is
\begin{align}
\mu_h =  \frac{7 p_T+  M_V}{6} \,.
\end{align}
In the comparison to data, we use the scales in Eq.~\eqref{scalevalues}, for consistency with~\cite{Becher:2011fc}, which used these scales in comparison to Tevatron data from run II.

\begin{figure}
\begin{center}
\hspace{-0.5cm}
\includegraphics[width=0.49\textwidth]{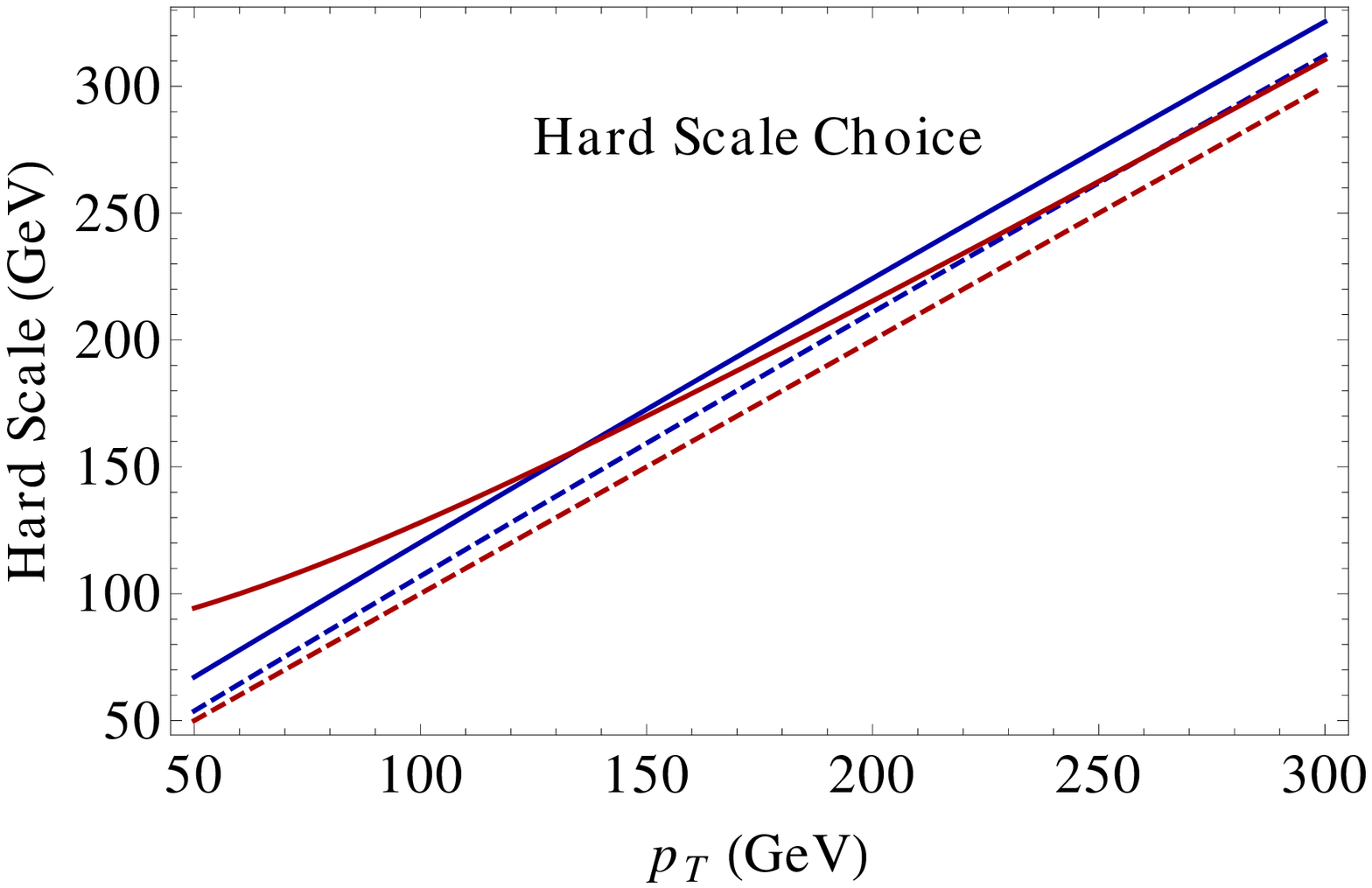}\hspace{0.2cm}
\includegraphics[width=0.49\textwidth]{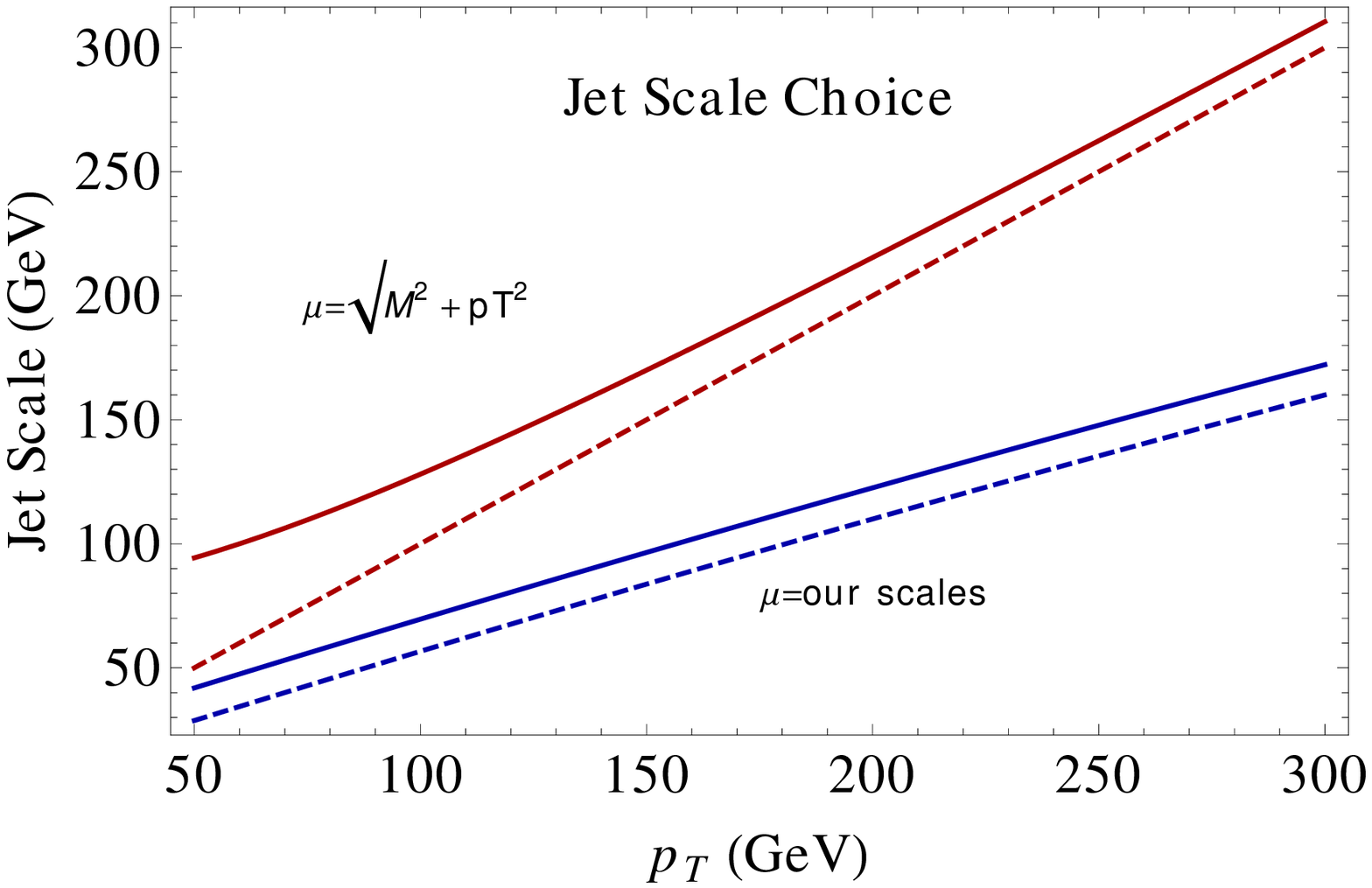}
\end{center}
\vspace{-0.4cm}
\caption{Comparison of our scale choices (blue) with the traditional choice $\mu = \sqrt{M_W^2 + p_T^2}$ (red)
for photon (dashed) and $W$ boson (solid). For the hard scale, there is not much difference. 
On the other hand, the natural jet (and soft) scales are lower than the traditional choice, but
higher than the fixed scale $\mu=M_W$. 
}
\label{fig:scalesMt}
\end{figure}

From Figures~\ref{fig:scalefig} or \ref{fig:scalesnlo}, it is obvious by eye that the natural jet scale is lower than the natural hard scale. While the hard scale is actually fairly close to a common scale choice $\mu = \sqrt{M_V^2+ p_T^2}$, as shown on the left of Figure~\ref{fig:scalesMt}, the jet scale is significantly lower. Both scales are higher than the fixed scale $\mu = M_V$ which was used in comparison to {\sc mcfm} in the {\sc atlas} study of the $W$ spectrum~\cite{Aad:2011fp}.

 To further emphasize the importance of scale choices, we show in Figure~\ref{fig:nlobands} the relative difference in the NLO prediction from the different parameterizations for the $W$ spectrum. The band
corresponds to a region $\frac{1}{2} \mu(p_T) < \mu < 2 \mu(p_T)$ where $\mu(p_T)$ is either
$M_W$, our hard scale, or the popular choice $\sqrt{M_W^2 + p_T^2}$. For $p_T \sim M_W$, all scales give
comparable results while for large $p_T$, the fixed scale gives a prediction significantly higher than either of the other two parameterizations. If we used the jet scale instead of the hard scale, the band would be closer to the $\mu=M_W$ band. Thus it is important to choose the appropriate scale in the appropriate place to get an accurate prediction.

\begin{figure}
\begin{center}
\includegraphics[width=0.7\textwidth]{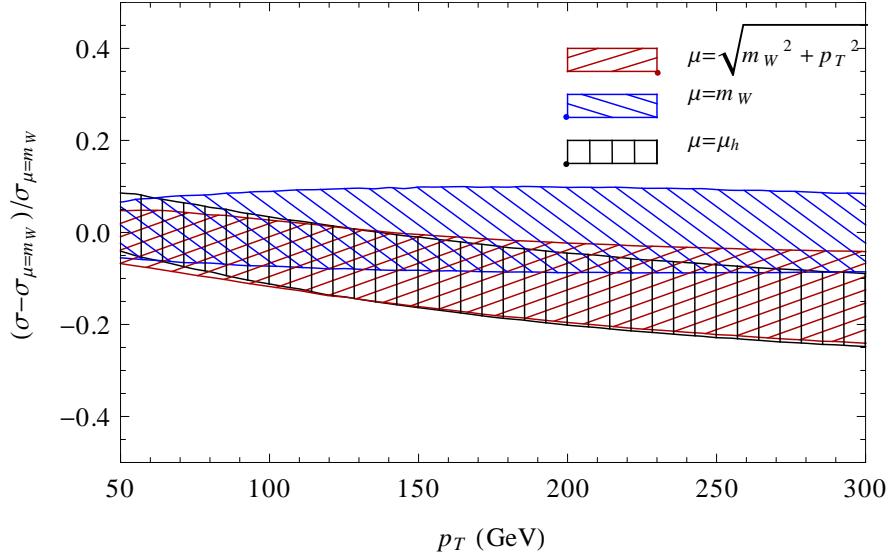}
\end{center}
\vspace{-0.4cm}
\caption{Scale variations at next-to-leading order. The blue southeast stripes show the scale variation 
of the NLO calculation (called NNLO in {\sc fewz}) with $\mu=\mu_f=\mu_r=M_W$, as in the {\sc atlas} paper. The
red northeast stripes show the prediction using $\mu_f=\mu_r=\sqrt{M_W^2 + p_T^2}$ and the 
black vertical stripes have $\mu_f$ and $\mu_r$ set to the scales in Eq.~\eqref{scalevalues}. Bands correspond to varying 
$\mu=\mu_f=\mu_r$ by factors of two from these default scales.}
\label{fig:nlobands}
\end{figure}

\begin{figure}
\begin{center}
\includegraphics[width=1.0\textwidth]{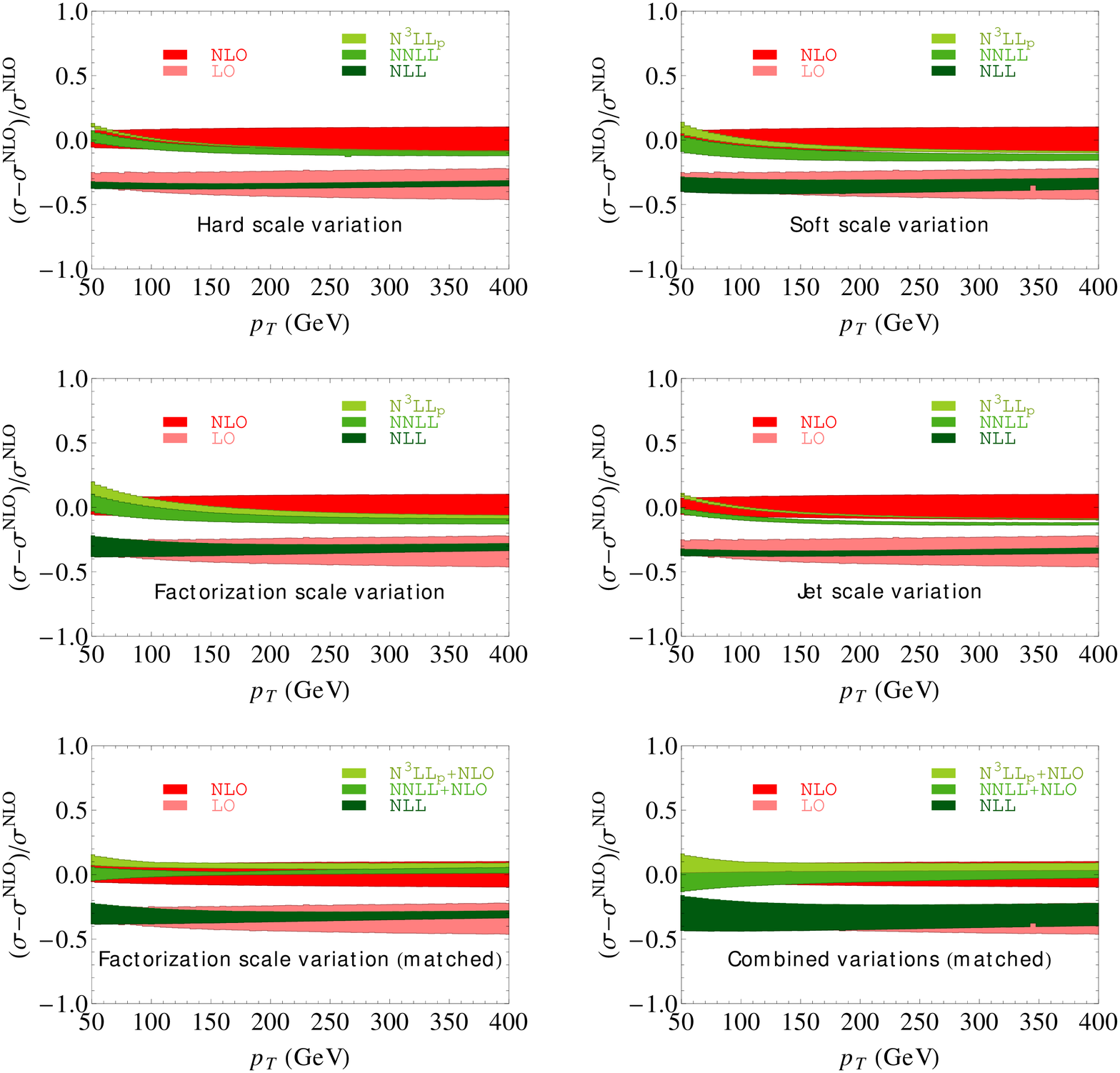}
\end{center}
 \vspace{-0.4cm}
 \caption{The  cross section for $W^+$ production at the LHC 7 TeV for various orders in perturbation 
theory, normalized relative to the next-to-leading fixed-order curves. In the top four panels the
 resummed curves are {\it not} matched to fixed order, which shows how including just the
 logarithms compares to the full result. Bands come from variations of the hard, jet, soft and factorization scales by factors of 2 around our default scales are shown and taking the maxima within that variation
 region. The fifth panel shows the factorization scale variation after matching. The sixth
panel is the uncertainty from adding the hard, jet, soft and factorization uncertainties in quadrature.
}
\label{fig:variations}
\end{figure}

Having determined the default values for the scales in Eq.~\eqref{scalevalues}, we can compute the resummed distribution. As discussed above, we include all ingredients for N${}^3$LL accuracy, except for the two-loop non-logarithmic terms in the hard, jet and soft functions. We match to NLO fixed order and denote our highest order resummed result by N${}^3$LL$_{\rm p}+$NLO, where the subscript ``p'' stands for partial.
Convergence in perturbation theory and the relative size of various scale variations are shown in Figure~\ref{fig:variations}. To generate the bands in this plot, we determine the maximum and minimum cross section obtained when varying each scale up and down by a factor of $2$ around its default value. In contrast to the fixed-order result, the scale dependence is not monotonic (cf. Figure~\ref{fig:scalefig}). To determine the maximum and minimum, we compute the cross section at $\frac{1}{2}$, $2$ and the central value, fit a parabola to those three points and take the maximum and minimum along the parabola. 

Curves in the first four panels of Figure~\ref{fig:variations} are not matched to fixed order. The relatively
large factorization scale uncertainty comes about because the $\muf
$ dependence is only canceled
in the resummed distribution near threshold. The full $\muf
$ dependence at NLO is removed once the theory is matched to the fixed-order distribution, as it is in the bottom two panels. The combined uncertainty that we use for our final error estimates is the quadratic sum of the hard, jet,
soft and factorization scale uncertainties:
\begin{equation}
\Delta \sigma = \sqrt{ (\Delta_h \sigma)^2 + (\Delta_j \sigma)^2+(\Delta_s \sigma)^2+(\Delta_f \sigma)^2}
\end{equation}
This is a conservative error estimate. We observe that the N${}^2$LL and N${}^3$LL$_{\rm p}$ scale variation bands overlap, but the increase in the cross section from NLL to N${}^2$LL is larger than the NLL band. The same behavior is also seen when going from the LO to the NLO fixed-order result. The corresponding bands would overlap had we evaluated the LO and NLL results with leading-order PDF sets which have a larger value of $\alpha_s$, instead of the NNLO PDFs we use throughout. The increase in the cross section from NLL to N${}^2$LL is mostly due to the one-loop constants in the soft and hard functions, as can be seen from the right panel of Figure~\ref{fig:scalefig}. We have checked how much of a shift the known two-loop jet and soft function constants induce and find that it is below a per cent.

\section{Comparison with LHC data}
We are now ready to compare to LHC data. We discuss separately the two processes we study, direct photon and $W$ production. For numerical work we use the NNLO MSTW 2008 PDF set and its associated
$\alpha_s(M_Z) = 0.1171$~\cite{Martin:2009iq}. We also use $M_W = 80.399$ GeV, $\alpha_{\rm e.m.} = 127.916^{-1}$,
$\sin^2\theta_W = 0.2226$, $V_{ud} = 0.97425$, $V_{us} = 0.22543$, $V_{ub} = 0.00354$, $V_{cd} = 0.22529$,
$V_{cs} = 0.97342$ and $V_{cb} = 0.04128$.

\subsection{Direct photon}
For direct photon production, to be consistent with the comparison to Tevatron data in ~\cite{Becher:2009th},
we use the scale choices from that paper
\begin{align}
\mu_h &= p_T \,,  \nonumber \\
\mu_j & = \frac{p_T}{2}\left (1-2 \frac{p_T}{\sqrt{s}}\right)  \,. \label{mujchoice} 
\end{align}
These are slightly simpler than those in Eq.~\eqref{scalevalues}, but equivalent within the uncertainties.

The direct photon spectrum is complicated by the requirement of photon isolation which is necessary to
remove hadronic backgrounds, such as $\pi^0$ decays. The {\sc atlas} study~\cite{Aad:2011tw}
 required energy in an cone of $R=0.4$ around the photon to have energy less than $4\,{\rm GeV}$. 
To include this effect rather than matching to the inclusive fixed-order NLO calculation, we match
to the NLO calculation with isolation and fragmentation contributions using the Monte Carlo program
{\sc jetphox}~\cite{Catani:2002ny}.

\begin{figure}
\begin{center}
\includegraphics[width=0.55\textwidth]{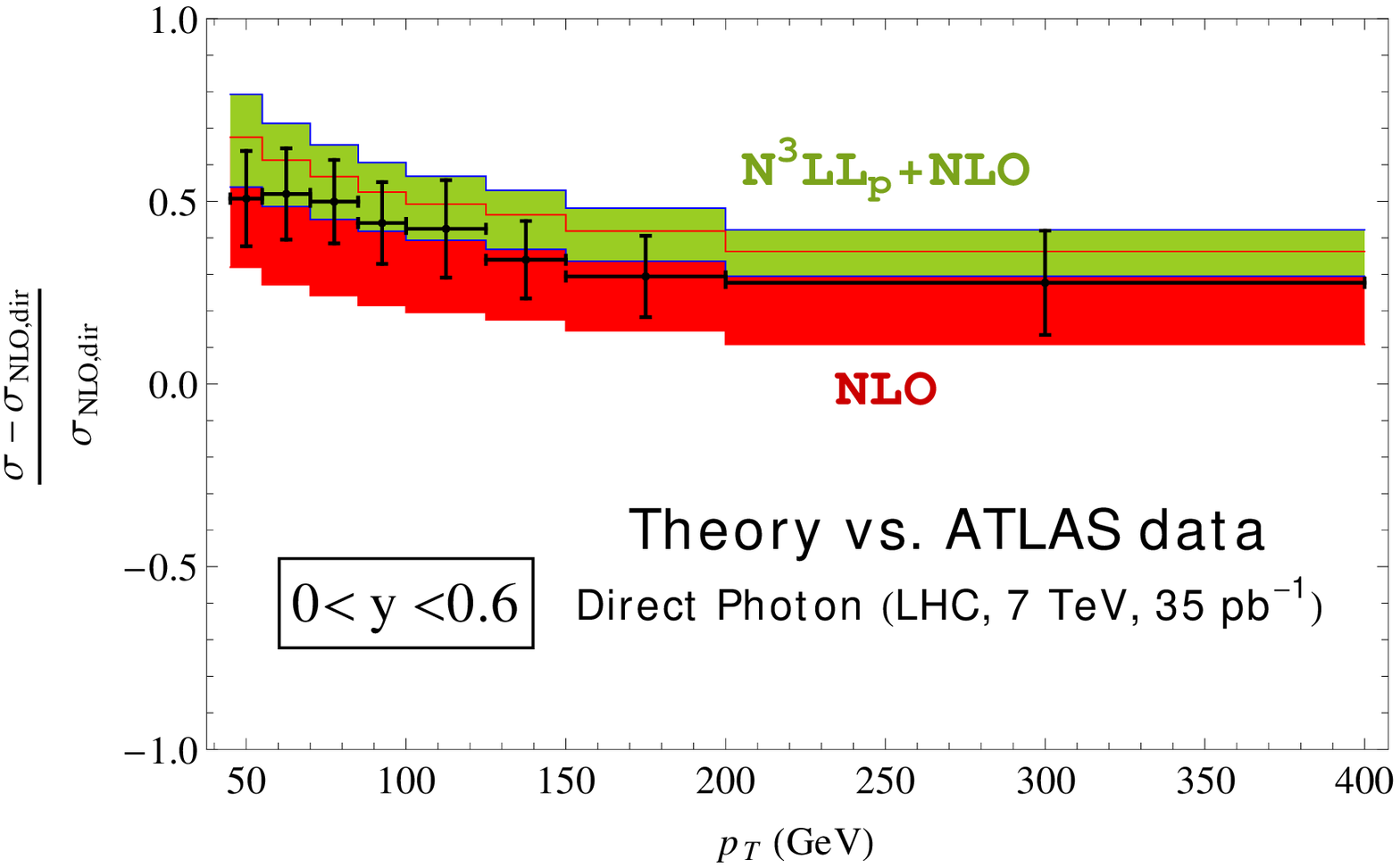}
\includegraphics[width=0.55\textwidth]{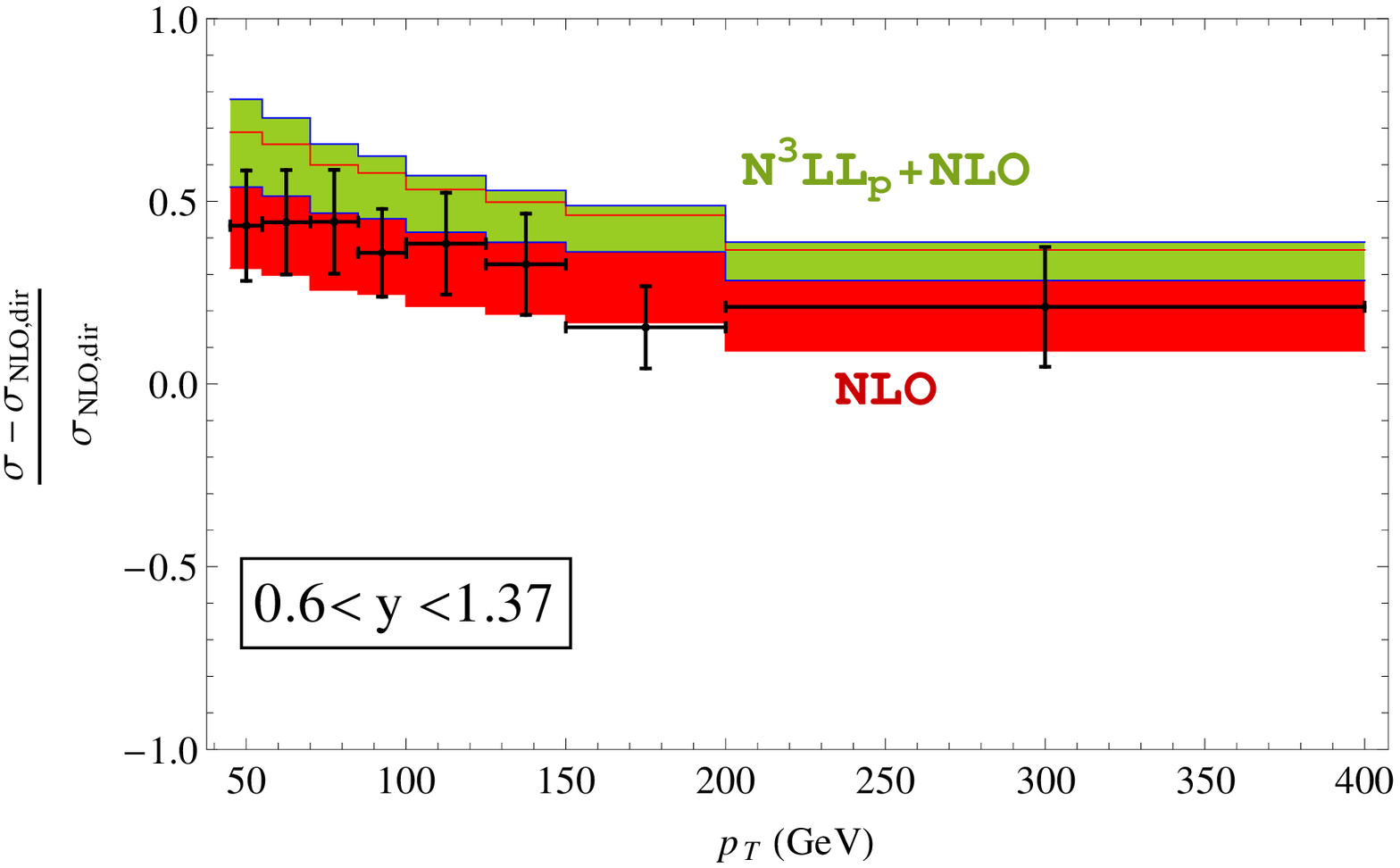}
\includegraphics[width=0.55\textwidth]{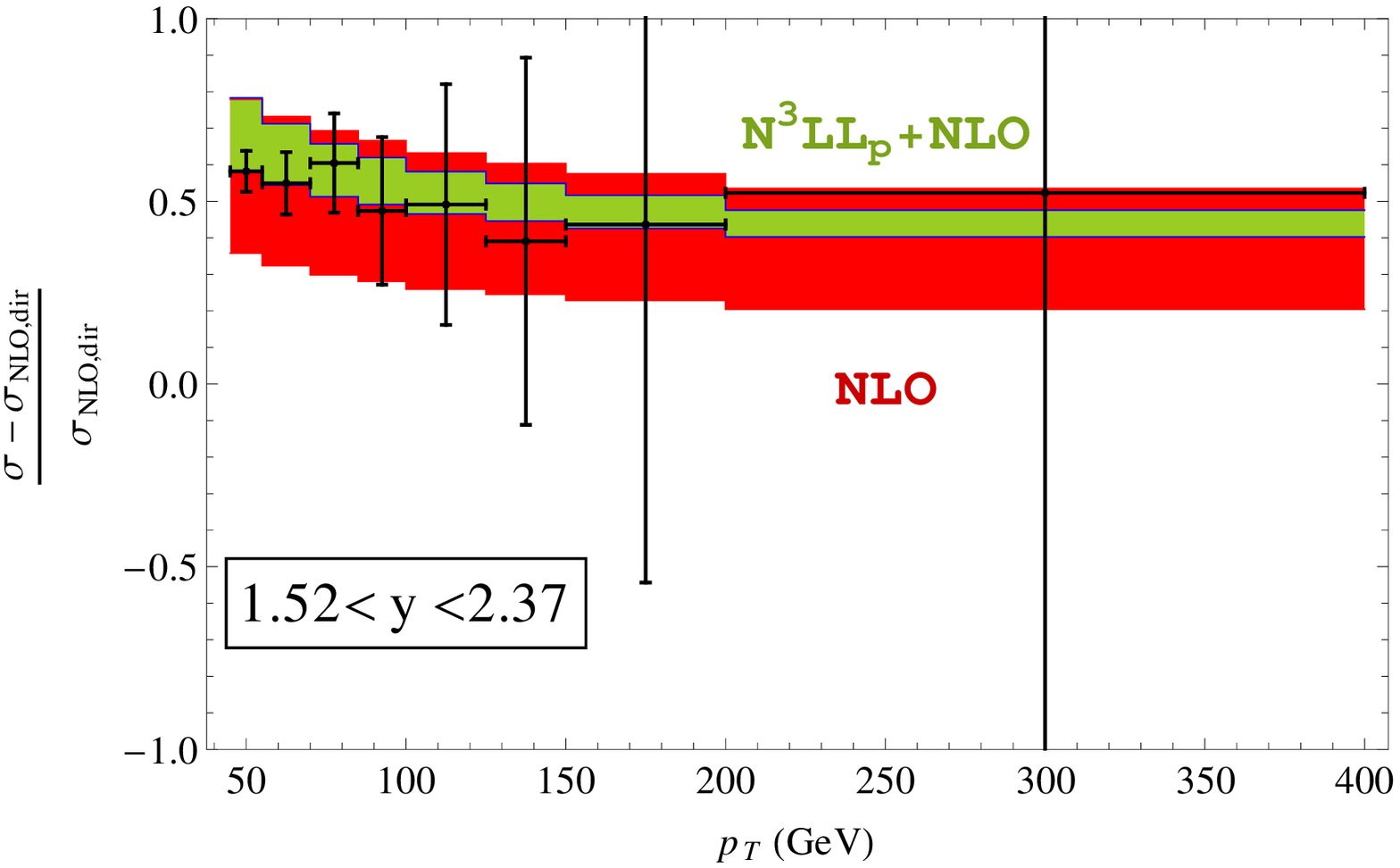}
\end{center}
\vspace{-0.4cm}
\caption{Comparison of theory to {\sc atlas} data for direct photon in various rapidity regions.
The normalization is such that 0 corresponds to the inclusive direct photon
spectrum at NLO without isolation or fragmentation. The NLO cross sections including 
fragmentation and isolation are slightly higher, especially at low $p_T$, since the enhancement
due to fragmentation outweighs the reduction due to isolation. Uncertainties on the NLO
correspond to factor of 2 variation of $\mu_f=\mu_r$ around its default value of $p_T$. For
the resummed curves, which also include isolation and fragmentation, uncertainties are hard, jet, soft and factorization variations by
factors of 2 added in quadrature. Uncertainties from fragmentation and isolation are {\it not} included,
and may be large.}
\label{fig:dpcomp}
\end{figure}

The {\sc atlas} data in~\cite{Aad:2011tw} includes 35 pb${}^{-1}$ of data separated into three rapidity
regions. The comparison of the theory prediction at NLO and at N${}^3$LL$_{\rm p}$+NLO order is shown in Figure~\ref{fig:dpcomp}. The theory and data are in agreement within uncertainties. It will
be interesting to update this comparison to a large data set, particularly if it includes
higher $p_T$.

\subsection{\boldmath $W$ boson}
The calculation of the $W$ boson $p_T$ spectrum at N${}^3$LL$_{\rm p}$+NLO order is significantly more challenging numerically than direct photon production, despite the identical factorization formulae. The extra scale, the boson mass, complicates the kinematics, which makes the integrals converge more slowly and the scale choices more complicated. Moreover, experimentally, since only the charged lepton from the $W$ decay is measured, acceptances
have to be included in comparing the inclusive $W$ spectrum to the measured distribution.

The acceptance cuts used by {\sc atlas} in~\cite{Aad:2011fp}
 were 
\begin{equation}
p_T^{l} > 20\, {\rm GeV}, \quad |\eta^{l}| < 2.4, \quad p_T^\nu > 25\, {\rm GeV}\,,
\end{equation}
and
\begin{equation}
m_T^W \equiv \sqrt{ ( |{\vec p_T}^l| +|{\vec p_T}^\nu|)^2 - |\vec{p_T}^l +\vec{p_T}^\nu|^2} \ge
 40\, {\rm GeV}.
\end{equation}
where $l$ is an electron or muon.

\begin{figure}
\begin{center}
\includegraphics[width=0.65\textwidth]{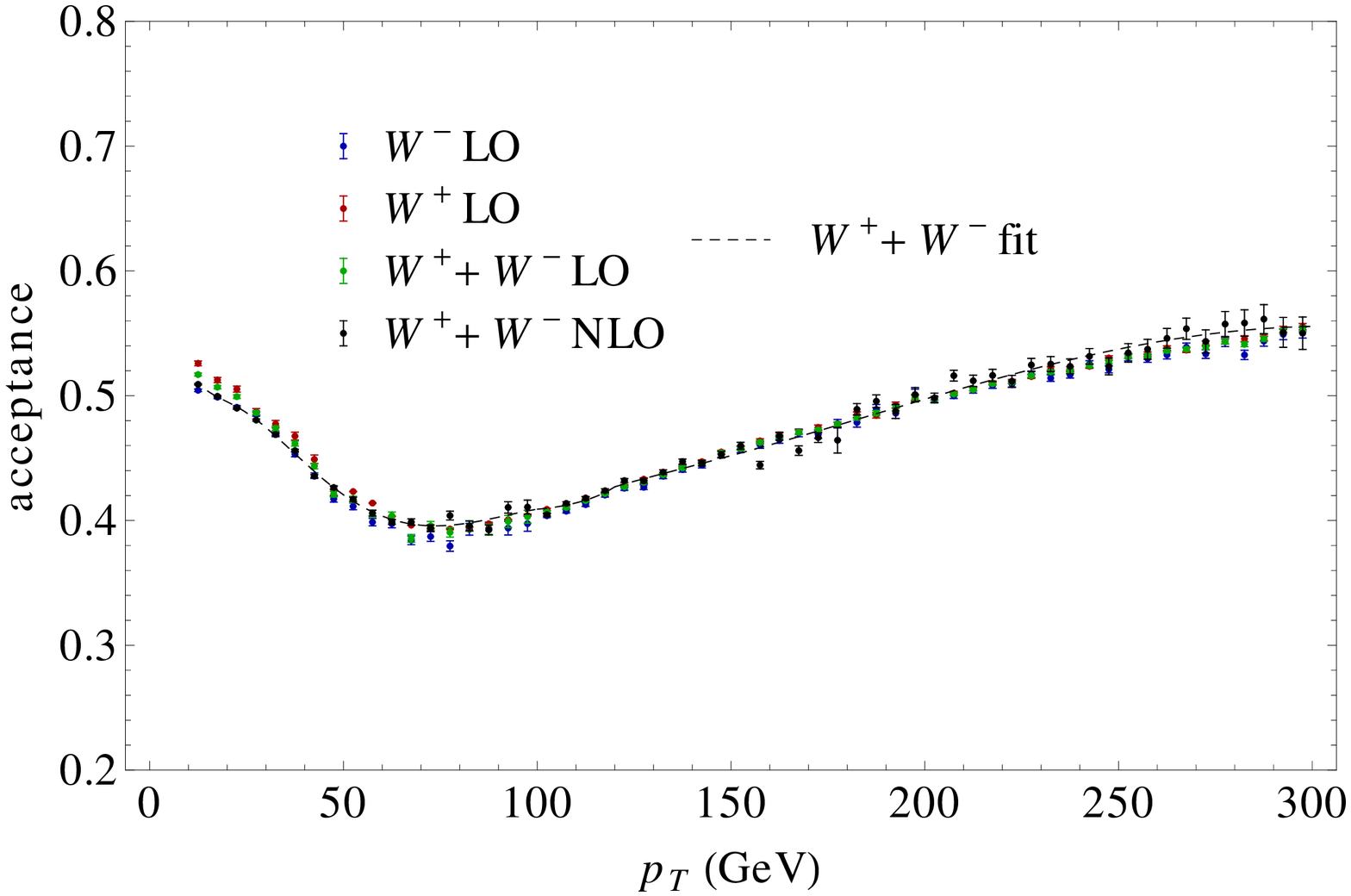}
\end{center}
\vspace{-0.4cm}
\caption{Kinematic acceptances according to {\sc atlas} cuts. There is little difference in the acceptances for $W^+$ or $W^-$ bosons. However, there is about a 2\% increase in the efficiency when going from leading-order  to next-to-leading order in perturbation theory (all computed with {\sc fewz}).
The NLO acceptances are used in our comparison with {\sc atlas} data.}
\label{fig:acceptances}
\end{figure}

To apply these cuts to our inclusive sample, we multiply our inclusive cross sections by acceptances,
given by the ratio of the inclusive $W$ cross section to the $W$ cross section in the fiducial volume as a function of $p_T$. These acceptances are shown in Figure~\ref{fig:acceptances}. We calculate them at LO and NLO for $W^+$ and $W^-$ using the program
{\sc fewz}~\cite{Gavin:2010az}. We find no significant difference between the leading order and next-to-leading order acceptances. For the numerical work, we use a smooth function fit to the inclusive NLO acceptances for $W^+ + W^-$, also shown in Figure~\ref{fig:acceptances}. 

The total cross sections we find at the  $\sqrt{s} = 7$ TeV LHC, using $\mu_f = \mu_r = M_W$ are
\begin{align}
 \sigma( p p\to W^+ \to \mu^+ \nu)_{\mathrm{inc}} &= (6204\pm0.7)\, {\mathrm{pb}}\,, \qquad
 \sigma( p p\to W^+\to \mu^+ \nu)_{\mathrm{fid}} = (3061 \pm 3.0)\, {\mathrm{pb}}\,, \nonumber\\
 \sigma( p p\to W^-\to \mu^- \bar{\nu})_{\mathrm{inc}} &= (4326\pm0.5)\, {\mathrm{pb}}\,, \qquad
 \sigma( p p\to W^-\to \mu^- \bar{\nu})_{\mathrm{fid}} = (2038\pm 1.9) \, {\mathrm{pb}}\,,
\end{align}
where ``inc'' refers to the inclusive cross section and ``fid'' to the fiducial cross section (with cuts). Errors
are integration errors from {\sc fewz}.
Dividing these by the branching ratio to muons, $BR(W\to \mu \nu) = 0.1083$ gives the total inclusive
cross section. To compare to data, we take our theoretical calculation of the inclusive
differential cross section, multiply by the acceptance curves and then 
divide by the total cross section in the fiducial volume, $47.04\,{\rm nb}$.  This lets us compare directly to the {\sc atlas} data, which is normalized to the total number of events in the fiducial volume.

\begin{figure}
\begin{center}
\includegraphics[width=0.8\textwidth]{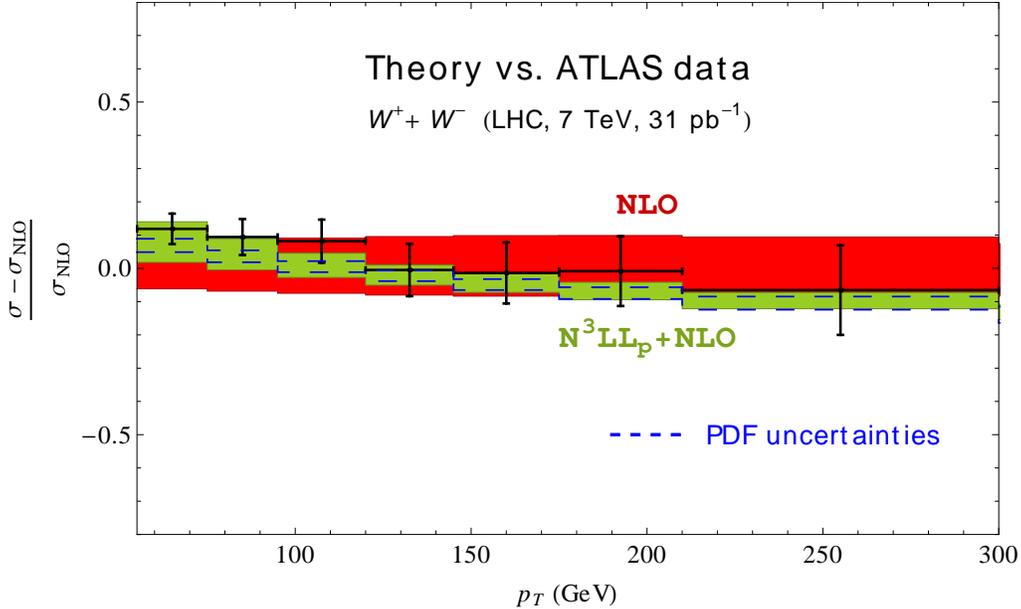}
\end{center}
\vspace{-0.4cm}
\caption{Comparison of theory to {\sc atlas} data for the $W$ spectra. The red band is the NLO prediction,
using $\mu_f = \mu_r = M_W$, as in~\cite{Aad:2011fp}. The N${}^3$LL${}_{\mathrm p}$ + NLO prediction, in green, is
in excellent agreement with the data. Dashed blue lines indicate PDF uncertainties which are of
order the scale uncertainties at N${}^3$LL${}_{\mathrm p}$ + NLO order.}
\label{fig:atlasWcompare}
\end{figure}

The comparison to the {\sc atlas} data is shown in Figure~\ref{fig:atlasWcompare}. The agreement is excellent. In this plot, results
are shown normalized to the NLO prediction with $\mu_f=\mu_r= M_W$. As we have argued in the previous section,
this is not a good scale choice in the large $p_T$ region.
We use $\mu=M_W$ as the basis for our comparison since it is the scale choice used by {\sc atlas} in~\cite{Aad:2011fp}, and, therefore, our NLO calculation can be directly compared to their calculation labeled
MCFM in Figure~7 of~\cite{Aad:2011fp}.

There are two important qualitative conclusions that can be drawn from Figure~\ref{fig:atlasWcompare}. The first
is that scale choices are important. Although the data is within the NLO uncertainty, comparing the NLO band in this
plot, which uses a fixed scale $\mu=M_W$ with those in Figure~\ref{fig:nlobands}, we see that the downward
trend in the data indicates a clear preference that $\mu$ should increase with $p_T$. However, as we have argued, there is no natural scale choice at fixed order, since multiple scales are present. Our procedure of determining these scales numerically is supported quantitatively by the 
agreement between the N${}^3$LL$_{\rm p}$+NLO band and the data in this figure.

The second qualitative conclusion concerns the PDF uncertainties. These are shown as the blue dashed lines in Figure~\ref{fig:atlasWcompare}. The PDF uncertainties are smaller than the uncertainties on the data and the NLO
scale uncertainties, but of the same order as the scale uncertainties of the resummed distribution. This indicates that PDF fits could be improved using the $W$ spectra, but only if resummation is included (or perhaps if the NNLO result becomes available).

\section{Conclusions}
In this paper, we have compared theoretical predictions for the direct photon and $W$ boson spectra at high $p_T$ to measurements performed by the {\sc atlas} collaboration using LHC data. The predictions were
performed using the exact cross section at NLO in $\alpha_s$ (the highest order known),
 supplemented with additional terms to all orders in $\alpha_s$ coming from a threshold expansion. These extra terms correspond to
large logarithms associated with infrared singularities of the recoiling jet. To isolate these terms,
the resummed calculation is performed near the partonic threshold, in which the $p_T$ of the vector boson
is maximal for the given value of the partonic center-of-mass energy. In this limit, the cross section factorizes and the logarithmic terms can then included to all orders in perturbation theory. These terms usually give the dominant contribution to the the cross section. In the photon case, the fragmentation cross section  and isolation corrections were added also, using the program {\sc jetphox}.

A main advantage of the resummed cross section, which was calculated using effective theory in~\cite{Becher:2009th} and~\cite{Becher:2011fc}, is that it has well-defined scales associated with different phase space regions. Unlike a fixed-order calculation, which merges all the scales into one, the effective field theory allows one to choose the scales appropriate for the relevant regions. We employed a numerical procedure to determine which scales are appropriate. This removes a source of uncertainty from the fixed-order calculation, namely, which value of the scale $\mu$ should be adopted. Typical fixed-order calculations choose scales like $M_W$, $p_T$ or $\sqrt{p_T^2 + M_W}$.  We find that a single such parametrization is insufficient: while the hard scale is naturally close to $\sqrt{p_T^2 + M_W}$, the jet and soft scales are naturally lower. 

The results of our comparison with data are shown in Figures~\ref{fig:dpcomp} and~\ref{fig:atlasWcompare}. The photon case is complicated by the requirement of photon isolation, while the $W$ case is complicated by the missing energy and the necessary acceptance cuts on the lepton.  We found good agreement for the direct photon case and excellent agreement for the case of the $W$ boson. In the $W$ boson case, one can clearly see the importance of proper scale choices. Moreover, the reduction of theoretical uncertainty when resummation is included is enough to make it comparable to the PDF uncertainty. These comparisons provide a  convincing demonstration of the relevance of resummation for LHC physics.

In the future, it would be useful to compare our prediction to the direct photon data available from {\sc cms} \cite{Chatrchyan:2011ue}. It would also be interesting to compare to Drell-Yan spectra from intermediate $Z$ bosons at high $p_T$\cite{Chatrchyan:2011wt,Aad:2011gj} and to direct photon and $W$ data at higher luminosity. Improvements of the theoretical description could be achieved by including the full two-loop hard, jet and soft functions and the fixed-order NNLO calculation, once it becomes available. Furthermore, on top of the QCD effects, one should also include electroweak Sudakov logarithms~\cite{Kuhn:2004em,Hollik:2007sq}, which will have a noticeable effect at high $p_T$. In addition, we hope to eventually provide a publicly available code to produce the resummed results.

\section{Acknowledgments}
The authors would like to thank F.~Bucci, A.~Hamilton and G.~Marchiori for discussions of the direct photon process and C.~Mills and J.~Guimares da Costa for discussions about the {\sc atlas} $W$ measurement. We thank R.~Gonsalves and F.~Petriello for help with the programs {\sc q${}_T$} and {\sc fewz}, respectively, and X.~Tormo for comments on the manuscript. The work of TB and CL is supported by the Swiss National Science Foundation (SNF) under grant 200020-140978 and the Innovations- und Kooperationsprojekt C-13 of the Schweizerische Universit\"atskonferenz (SUK/CRUS). MDS is supported by the US Department of Energy under grant DE-SC003916. 

\begin{appendix}


\section{One-loop hard function}
In this appendix we give the result for the hard function in both the annihilation and the Compton channel. The functions are related by crossing symmetry, but the analytic continuation from one channel to the other is not entirely trivial. The hard function is obtained from the result (A.9) in \cite{Arnold:1988dp} after performing renormalization. 

For the annihilation channel, we have
\begin{align}
H_\qq(u,t) &= 1 + \frac{\alpha_s}{4\pi} \left\{ C_A \frac{\pi^2}{6} + C_F \left(-16 + \frac{7\pi^2}{3} \right)
+ 2 C_A \ln^2\frac{s}{M_V^2} + C_A \ln^2\frac{M_V^2 - t}{M_V^2} \right.
\nonumber
\\
& \left.  + C_A \ln^2\frac{M_V^2 - u}{M_V^2} + \ln\frac{s}{M_V^2} \left(-6 C_F - 2 C_A \ln\frac{s^2}{t u}\right)
- C_A \ln^2\frac{t u}{M_V^4} - 6 C_F \ln\frac{\mu^2}{s} \right.
\nonumber
\\
& \left. - 2 C_A \ln\frac{s^2}{t u} \ln\frac{\mu^2}{s}
+ \left(-C_A - 2 C_F\right) \ln^2\frac{\mu^2}{s} \right.
\nonumber
\\
& \left. + 2 C_A \Li2\left(\frac{M_V^2}{M_V^2 - t}\right)
+ 2 C_A \Li2\left(\frac{M_V^2}{M_V^2 - u}\right)
\right\}
\nonumber
\\
& + \frac{\alpha_s}{4\pi} \frac{2}{T_0(u,t)} \left\{
C_F \left( \frac{s}{s + t} + \frac{s + t}{u}+ \frac{s}{s + u} + \frac{s + u}{t} \right) \right.
\nonumber
\\
& \left.+ \left(-C_A + 2 C_F\right) \left[-\frac{M_V^2 \left( t^2 + u^2 \right)}{t u \left(t + u\right)}
+ 2 \left(\frac{s^2}{\left(t + u\right)^2} + \frac{2 s}{t + u}\right) \ln\frac{s}{M_V^2}\right] \right.
\nonumber
\\
& \left. + \left( C_A\frac{t}{s + u}
+ C_F\frac{4 s^2 + 2 s t + 4 s u + t u}{\left(s + u\right)^2} \right) \ln\frac{-t}{M_V^2} \right.
\nonumber
\\
& \left. + \left( C_A\frac{u}{s + t}
+ C_F \frac{4 s^2 + 4 s t + 2 s u + t u}{\left(s + t\right)^2} \right) \ln\frac{-u}{M_V^2} \right.
\nonumber
\\
& \left. - \left(-C_A + 2 C_F\right)
\left[ \frac{s^2 + \left(s + u\right)^2}{t u} \left(\frac{1}{2} \ln^2\frac{s}{M_V^2} - \frac{1}{2} \ln^2\frac{M_V^2 - t}{M_V^2}
+ \ln\frac{s}{M_V^2} \ln\frac{-t}{s-M_V^2 } \right.\right.\right.
\nonumber
\\
& \left.\left.\left. + \Li2\left(\frac{M_V^2}{s}\right)
- \Li2\left(\frac{M_V^2}{M_V^2 - t}\right) \right) \right.\right.
\nonumber
\\
& \left.\left. + \frac{s^2 + \left(s + t\right)^2}{t u} \left( \frac{1}{2} \ln^2\frac{s}{M_V^2} - \frac{1}{2} \ln^2\frac{M_V^2 - u}{M_V^2}
+ \ln\frac{s}{M_V^2} \ln\frac{-u}{s-M_V^2} \right.\right.\right.
\nonumber
\\
& \left.\left.\left. + \Li2\left(\frac{M_V^2}{s}\right)
- \Li2\left(\frac{M_V^2}{M_V^2 - u}\right)
\right)
\right]
\right\}\, ,
\end{align}
and the Compton channel result reads
\begin{align}
H_{qg}(u,t) &= 1 + \frac{\alpha_s}{4\pi} \left\{
C_A\frac{7 \pi^2}{6} + C_F \left(-16 + \frac{\pi^2}{3}\right) - 6 C_F \ln\frac{s}{M_V^2}
-C_A \ln^2\frac{-s t}{M_V^4} + C_A \ln^2\frac{M_V^2-t}{M_V^2} \right.
\nonumber
\\
& \left. + 2 C_A \ln\frac{\left(s-M_V^2 \right) t}{M_V^2 u} \ln\frac{-u}{M_V^2}
+ C_A \ln^2\frac{-u}{M_V^2} - 2 C_A \ln\frac{\left(M_V^2 - s\right) s t}{M_V^2 u^2} \ln\frac{-u}{s} \right.
\nonumber
\\
& \left. - 2 C_A \ln^2\frac{-u}{s} - 2 C_F \ln^2\frac{-u}{s}
+ \left(-6 C_F + 2 C_A \ln\frac{t}{u} + 4 C_F \ln\frac{-u}{s} \right) \ln\frac{\mu^2}{s} \right.
\nonumber
\\
& \left. - \left(C_A + 2 C_F\right) \ln^2\frac{\mu^2}{s}
- 2 C_A \Li2\left(\frac{M_V^2}{s}\right) + 2 C_A \Li2\left(\frac{M_V^2}{M_V^2 - t}\right)
\right\}
\nonumber
\\
& + \frac{\alpha_s}{4\pi} \frac{2}{T_0(s,t)} \left\{
C_F \left(\frac{u}{s + u} + \frac{s + u}{t} + \frac{u}{t + u} + \frac{t + u}{s} \right) \right.
\nonumber
\\
& \left. + \left( C_A \frac{s}{t + u} + C_F \frac{s t + 2 s u + 4 t u + 4 u^2}{\left(t + u\right)^2}
\right) \ln\frac{s}{M_V^2} \right.
\nonumber
\\
&  \left. + \left(C_A \frac{t}{s + u} + C_F \frac{s t + 4 s u + 2 t u + 4 u^2}{\left(s + u\right)^2} \right) \ln\frac{-t}{M_V^2} \right.
\nonumber
\\
& \left. + \left(-C_A + 2 C_F\right) \left[-\frac{M_V^2 \left(s^2 + t^2\right)}{s t \left(s + t\right)}
+ 2 \left(\frac{2 u}{s + t} + \frac{u^2}{\left(s + t\right)^2}\right) \ln\frac{-u}{M_V^2} \right] \right.
\nonumber
\\
& \left. - \left(-C_A + 2 C_F\right) \left[ \frac{u^2 + \left(t + u\right)^2}{s t}
\left(\frac{1}{2} \ln^2\frac{s}{M_V^2}
- \frac{1}{2} \ln^2\frac{M_V^2 - u}{M_V^2} + \ln\frac{s}{M_V^2} \ln\frac{-u}{s-M_V^2}
\right.\right.\right.
\nonumber
 \\
 & \left.\left.\left. + \Li2\left(\frac{M_V^2}{s}\right) - \Li2\left(\frac{M_V^2}{M_V^2 - u}\right)
\right) \right.\right.
\nonumber
\\
& \left.\left. + \frac{u^2 + \left(s + u\right)^2}{s t}  \left(-\frac{\pi^2}{2}
- \frac{1}{2} \ln^2\frac{M_V^2 - t}{M_V^2} - \frac{1}{2} \ln^2\frac{M_V^2 - u}{M_V^2}
+ \ln\frac{-t}{M_V^2} \ln\frac{-u}{M_V^2} \right.\right.\right.
\nonumber
\\
& \left.\left.\left.  - \Li2\left(\frac{M_V^2}{M_V^2 - t}\right)
- \Li2\left(\frac{M_V^2}{M_V^2 - u}\right)
\right)
\right]
\right\}\,.
\end{align}
 
\end{appendix}


\begin{thebibliography}{1}



\bibitem{Aad:2011tw} 
  G.~Aad {\it et al.}  [ATLAS Collaboration],
  Phys.\ Lett.\ B {\bf 706}, 150 (2011)
  [arXiv:1108.0253 [hep-ex]].

\bibitem{Aad:2011fp} 
  G.~Aad {\it et al.}  [ATLAS Collaboration],
  Phys.\ Rev.\ D {\bf 85}, 012005 (2012)
  [arXiv:1108.6308 [hep-ex]].


\bibitem{Aurenche:1983ws}
P.~Aurenche, A.~Douiri, R.~Baier, M.~Fontannaz and D.~Schiff,
Phys.\ Lett.\  B {\bf 140}, 87 (1984).

\bibitem{Aurenche:1987fs}
P.~Aurenche, R.~Baier, M.~Fontannaz and D.~Schiff,
Nucl.\ Phys.\  B {\bf 297}, 661 (1988).

\bibitem{Gordon:1993qc}
L.~E.~Gordon and W.~Vogelsang,
Phys.\ Rev.\  D {\bf 48}, 3136 (1993).
  
\bibitem{Ellis:1981hk}
  R.~K.~Ellis, G.~Martinelli, R.~Petronzio,
  Nucl.\ Phys.\  {\bf B211}, 106 (1983).

\bibitem{Arnold:1988dp}
  P.~B.~Arnold, M.~H.~Reno,
  Nucl.\ Phys.\  {\bf B319}, 37 (1989).

\bibitem{Gonsalves:1989ar}
  R.~J.~Gonsalves, J.~Pawlowski, C.~-F.~Wai,
  Phys.\ Rev.\  {\bf D40}, 2245 (1989).
  
   \bibitem{qt}
R.~Gonsalves, http://www.physics.buffalo.edu/gonsalves/.


  
 \bibitem{mcfm}
J.~Campbell, K.~Ellis, C.~Williams, http://mcfm.fnal.gov/.



\bibitem{Melnikov:2006kv} 
  K.~Melnikov and F.~Petriello,
  Phys.\ Rev.\ D {\bf 74}, 114017 (2006)
  [hep-ph/0609070].

\bibitem{Gavin:2010az} 
  R.~Gavin, Y.~Li, F.~Petriello and S.~Quackenbush,
  Comput.\ Phys.\ Commun.\  {\bf 182}, 2388 (2011)
  [arXiv:1011.3540 [hep-ph]].
  
\bibitem{Catani:2009sm} 
  S.~Catani, L.~Cieri, G.~Ferrera, D.~de Florian and M.~Grazzini,
  Phys.\ Rev.\ Lett.\  {\bf 103}, 082001 (2009)
  [arXiv:0903.2120 [hep-ph]].
  
\bibitem{Balazs:1997xd} 
  C.~Balazs and C.~P.~Yuan,
  Phys.\ Rev.\ D {\bf 56}, 5558 (1997)
  [hep-ph/9704258].

\bibitem{Bozzi:2010xn} 
  G.~Bozzi, S.~Catani, G.~Ferrera, D.~de Florian and M.~Grazzini,
  Phys.\ Lett.\ B {\bf 696}, 207 (2011)
  [arXiv:1007.2351 [hep-ph]].
  
  
\bibitem{Becher:2010tm} 
  T.~Becher and M.~Neubert,
  Eur.\ Phys.\ J.\ C {\bf 71}, 1665 (2011)
  [arXiv:1007.4005 [hep-ph]].

\bibitem{Laenen:1998qw} 
  E.~Laenen, G.~Oderda and G.~F.~Sterman,
  Phys.\ Lett.\ B {\bf 438}, 173 (1998)
  [hep-ph/9806467].
    
\bibitem{Becher:2009th}
  T.~Becher, M.~D.~Schwartz,
  JHEP {\bf 1002}, 040 (2010).

\bibitem{Becher:2011fc} 
  T.~Becher, C.~Lorentzen and M.~D.~Schwartz,
  Phys.\ Rev.\ Lett.\  {\bf 108}, 012001 (2012)
  [arXiv:1106.4310 [hep-ph]].

  
  \bibitem{Bauer:2000yr}
 C.~W.~Bauer, S.~Fleming, D.~Pirjol and I.~W.~Stewart,
 Phys.\ Rev.\ D {\bf 63}, 114020 (2001).

\bibitem{Bauer:2001yt}
 C.~W.~Bauer, D.~Pirjol and I.~W.~Stewart,
 Phys.\ Rev.\ D {\bf 65}, 054022 (2002)

\bibitem{Beneke:2002ph}
 M.~Beneke, A.~P.~Chapovsky, M.~Diehl and T.~Feldmann,
 Nucl.\ Phys.\ B {\bf 643}, 431 (2002)
 
\bibitem{Kidonakis:1999ur}
  N.~Kidonakis, V.~Del Duca,
  Phys.\ Lett.\  {\bf B480}, 87-96 (2000).
  
\bibitem{Kidonakis:2003xm}
  N.~Kidonakis, A.~Sabio Vera,
  JHEP {\bf 0402}, 027 (2004).

 
\bibitem{Gonsalves:2005ng}
  R.~J.~Gonsalves, N.~Kidonakis, A.~Sabio Vera,
  Phys.\ Rev.\ Lett.\  {\bf 95}, 222001 (2005).

\bibitem{Kidonakis:2012su} 
  N.~Kidonakis and R.~J.~Gonsalves,
  arXiv:1201.5265 [hep-ph].
  
\bibitem{Kidonakis:2011hm} 
  N.~Kidonakis and R.~J.~Gonsalves,
  arXiv:1109.2817 [hep-ph].

\bibitem{Appell:1988ie} 
  D.~Appell, G.~F.~Sterman and P.~B.~Mackenzie,
  Nucl.\ Phys.\ B {\bf 309}, 259 (1988).

\bibitem{Catani:1998tm} 
  S.~Catani, M.~L.~Mangano and P.~Nason,
  JHEP {\bf 9807}, 024 (1998)
  [hep-ph/9806484].

\bibitem{Becher:2007ty}
  T.~Becher, M.~Neubert, G.~Xu,
  JHEP {\bf 0807}, 030 (2008).

\bibitem{Becher:2006nr}
  T.~Becher, M.~Neubert,
  Phys.\ Rev.\ Lett.\  {\bf 97}, 082001 (2006).
  [hep-ph/0605050].
  
\bibitem{Becher:2006mr} 
  T.~Becher, M.~Neubert and B.~D.~Pecjak,
  JHEP {\bf 0701}, 076 (2007)
  [hep-ph/0607228].


\bibitem{Becher:2006qw}
  T.~Becher, M.~Neubert,
  Phys.\ Lett.\  {\bf B637}, 251-259 (2006).
  [hep-ph/0603140].

\bibitem{Becher:2010pd}
  T.~Becher, G.~Bell,
  Phys.\ Lett.\  {\bf B695}, 252-258 (2011).

\bibitem{Becher:2012za} 
  T.~Becher, G.~Bell and S.~Marti,
  JHEP {\bf 1204}, 034 (2012)
  [arXiv:1201.5572 [hep-ph]].
  

\bibitem{De Fazio:1999sv}
F.~De Fazio and M.~Neubert,
JHEP {\bf 9906}, 017 (1999)
[arXiv:hep-ph/9905351].

\bibitem{Becher:2009qa} 
  T.~Becher and M.~Neubert,
  JHEP {\bf 0906}, 081 (2009)
  [arXiv:0903.1126 [hep-ph]].

  
\bibitem{Garland:2002ak} 
  L.~W.~Garland, T.~Gehrmann, E.~W.~N.~Glover, A.~Koukoutsakis and E.~Remiddi,
  Nucl.\ Phys.\ B {\bf 642}, 227 (2002)
  [hep-ph/0206067].
  
\bibitem{Gehrmann:2011ab} 
  T.~Gehrmann and L.~Tancredi,
  JHEP {\bf 1202}, 004 (2012)
  [arXiv:1112.1531 [hep-ph]].

\bibitem{Ahrens:2008nc} 
  V.~Ahrens, T.~Becher, M.~Neubert and L.~L.~Yang,
  Eur.\ Phys.\ J.\ C {\bf 62}, 333 (2009)
  [arXiv:0809.4283 [hep-ph]].
  

  
\bibitem{Schwartz:2007ib} 
  M.~D.~Schwartz,
  Phys.\ Rev.\ D {\bf 77}, 014026 (2008)
  [arXiv:0709.2709 [hep-ph]].

\bibitem{Becher:2008cf} 
  T.~Becher and M.~D.~Schwartz,
  JHEP {\bf 0807}, 034 (2008)
  [arXiv:0803.0342 [hep-ph]].

\bibitem{Chien:2010kc} 
  Y.~-T.~Chien and M.~D.~Schwartz,
  JHEP {\bf 1008}, 058 (2010)
  [arXiv:1005.1644 [hep-ph]].
\bibitem{Feige:2012vc} 
  I.~Feige, M.~Schwartz, I.~Stewart and J.~Thaler,
  arXiv:1204.3898 [hep-ph].


\bibitem{Martin:2009iq}
  A.~D.~Martin, W.~J.~Stirling, R.~S.~Thorne, G.~Watt,
  Eur.\ Phys.\ J.\  {\bf C63}, 189-285 (2009).
  
  \bibitem{Catani:2002ny} 
  S.~Catani, M.~Fontannaz, J.~P.~Guillet and E.~Pilon,
  JHEP {\bf 0205}, 028 (2002)
  [hep-ph/0204023].
  
\bibitem{Chatrchyan:2011ue} 
  S.~Chatrchyan {\it et al.}  [CMS Collaboration],
  Phys.\ Rev.\ D {\bf 84}, 052011 (2011)
  [arXiv:1108.2044 [hep-ex]].
  
\bibitem{Chatrchyan:2011wt} 
  S.~Chatrchyan {\it et al.}  [CMS Collaboration],
  Phys.\ Rev.\ D {\bf 85}, 032002 (2012)
  [arXiv:1110.4973 [hep-ex]].
\bibitem{Aad:2011gj} 
  G.~Aad {\it et al.}  [ATLAS Collaboration],
  Phys.\ Lett.\ B {\bf 705}, 415 (2011)
  [arXiv:1107.2381 [hep-ex]].

  
  \bibitem{Kuhn:2004em}
  J.~H.~K\"uhn, A.~Kulesza, S.~Pozzorini, M.~Schulze,
  Phys.\ Lett.\  {\bf B609}, 277-285 (2005); %
  Nucl.\ Phys.\  {\bf B727}, 368-394 (2005); %
  Phys.\ Lett.\  {\bf B651}, 160-165 (2007); %
  Nucl.\ Phys.\  {\bf B797}, 27-77 (2008).
  
\bibitem{Hollik:2007sq}
  W.~Hollik, T.~Kasprzik, B.~A.~Kniehl,
  Nucl.\ Phys.\  {\bf B790 } (2008)  138-159.



  
\end{thebibliography}
\end{document}